\documentclass[12pt,preprint]{aastex}

\usepackage{graphicx}
\usepackage{hyperref}

\def\Msun{M_{\odot}}

\begin{document}

\shorttitle{Strong Radiative Feedback In Kinetic Mode}
\title {AGN feedback in an isolated elliptical galaxy: the effect of strong radiative feedback in the kinetic mode}
\author{Zhaoming Gan, Feng Yuan
\affil{Key Laboratory for Research in Galaxies and Cosmology, Shanghai Astronomical Observatory, Chinese Academy of Sciences, 80 Nandan Road, Shanghai 200030, China; fyuan@shao.ac.cn}
Jeremiah P. Ostriker
\affil{Department of Astrophysical Sciences, Princeton University, Princeton, NJ 08544, USA; Department of Astronomy, Columbia University, NYC, NY 10027, USA}
Luca Ciotti
\affil{Department of Physics and Astronomy, University of Bologna, via Ranzani 1, 40127 Bologna, Italy}
and Gregory S. Novak
\affil{Observatoire de Paris, LERMA, CNRS, 61 Av de l'Observatoire, 75014 Paris, France} }


\begin{abstract}
Based on two-dimensional high resolution hydrodynamic numerical simulation, we study the mechanical and radiative feedback effects from the central AGN on the cosmological evolution of an isolated elliptical galaxy. Physical processes such as star formation and supernovae are considered.
The inner boundary of the simulation domain is carefully chosen so that the fiducial Bondi radius is resolved and the accretion rate of the black hole is determined self-consistently.
In analogy to previous works, we assume that the specific angular momentum of the galaxy is low.
It is well-known that when the accretion rates are high and low, the central AGNs will be in cold and hot accretion modes, which correspond to the radiative and kinetic feedback modes, respectively. The emitted spectrum from the hot accretion flows is harder than that from the cold accretion flows,  which could result in a higher Compton temperature accompanied by a more efficient radiative heating, according to previous theoretical works. Such a difference of the Compton temperature between the two feedback modes, the focus of this study, has been neglected in previous works. Significant differences in the kinetic feedback mode are found as a result of the stronger Compton heating and accretion becomes more chaotic. More importantly,
if we constrain models to correctly predict black hole growth and AGN duty cycle after cosmological evolution, we find that the favored model parameters are constrained: mechanical feedback efficiency
diminishes with decreasing luminosity (the maximum efficiency being $\simeq 10^{-3.5}$) and X-ray Compton temperature increases with decreasing luminosity{, although models with fixed mechanical efficiency and Compton temperature can be found that are satisfactory as well.}
{We conclude that radiative feedback in the kinetic mode is much more important than previously thought}.
\end{abstract}

\keywords{Active galactic nucleus; Elliptical galaxy; Feedback; Galaxy evolution; Radiation}

\maketitle


\section{Introduction} \label{sec:introduction}
There is accumulating evidence showing that the evolution of host galaxies is tightly related to their central supermassive black holes (SMBHs) \citep{Fabian2012, Kormendy2013}. The most linkages are the strong correlations between the mass of SMBH and properties of the galactic spheroid, including luminosity \citep{Kormendy1995}, stellar velocity dispersion \citep{Gebhardt2000, Ferrarese2000, Tremaine2002, Gueltekin2009, Graham2011} and stellar mass \citep{Magorrian1998, Marconi2003, Haering2004}.
The general consensus is that powerful feedback from the central active galactic nuclei (AGNs) should play an important role in the formation and evolution processes of their host galaxies. Simple energetic arguments indicate that AGNs should be capable of heating the inner regions of galaxies to offset radiative cooling and further regulate the black hole growth \citep[e.g.,][]{Ciotti1997,Binney2001,Fabian2012}. Many
authors have proposed and performed successful theoretical models and numerical simulations to investigate how the feedback plays the role. However, because the length scales associated with AGN activity are so tiny compared to their host galaxy, the feedback process must operate over a huge dynamical range. This implies that the detailed AGN feedback mechanisms must be diverse and explains why this process is still incompletely understood \citep[for reviews, see][]{Ostriker2005,Peterson2006, McNamara2007, Cattaneo-Best2009, Fabian2012, Yuan-Narayan2014}.

To study AGN feedback, ideally, we should cover the whole range of lengths and timescales, from black hole event horizon to the galactic scales. Obviously, such a huge dynamical range is technically almost impossible to achieve by current 2- and 3-dimensional numerical simulations. In reality, different works focus on feedback at different scales. At the smallest black hole accretion scale, works have been done on outflow production driven by line force \citep[e.g.,][]{Kurosawa-Proga2009a, Kurosawa-Proga2009b, Liu2013a}, and on the radiative feedback due to the global Compton scattering \citep{Park-Ostriker1999, Park-Ostriker2001, YXO2009}. The latter effect has been invoked to explain the intermittent behavior of the black hole activity of some compact young radio sources \citep{Yuan-Li2011}. On the other hand, many works \citep[e.g.][]{DiMatteo2005,Johansson2009} have focused on the much larger galactic scale, from $\simeq 100$ pc to tens of kpc and the timescale from a fraction of Myr to several Gyr. Brighenti \& Mathews (e.g., \citeyear{Brighenti2002}; \citeyear{Brighenti2003}, see also \citeauthor{Mathews2003} \citeyear{Mathews2003}) investigated the AGN heating on the ISM in details, and found that the thermal feedback could break the cool-core structures which are observed in many clusters, i.e., the over-heating problem. Gaspari et al. (e.g., \citeyear{Gaspari2012}; \citeyear{Gaspari2013}) performed detailed simulations on the mechanical feedback, and they found that the mechanical feedback is favoured to solve the cooling flow problem, to explain to cold rims, and meantime to avoid over-heating.

Many works on the AGN feedback have been done in the context of isolated elliptical galaxies by Ciotti, Ostriker and their collaborators, both in one dimension
\citep{Ciotti1997,Ciotti2001,Ciotti2007,Ciotti2009b,Shin2010,Ostriker2010,Jiang2010} and two dimensions \citep{Novak2011,Novak2012}. In these works, the inner boundary is small enough to resolve the black hole accretion flow, typically a few pc. This is important since it can compile the accretion rate in a physical way, which is crucial to evaluate the effect of AGN feedback \citep{Novak2011}. When the Bondi radius \citep{Bondi1952} is not resolved, other more approximate methods of estimating the accretion rate \citep[e.g., see][]{Springel2005} must be utilized. The outer boundary is large enough to reach hundreds of kpc, i.e, the galactic outskirts. The timescale also covers a large range, from accretion timescale to galaxy evolution timescale. The interaction of the radiation and winds from the inner AGNs with the galactic gas and their effects on regulating the accretion are considered, together with physical processes such as supernovae heating and star formation. Overall, the above mentioned works evidenced that the most satisfactory models are the combined models with both mechanical feedback and radiative feedback,
in which mechanical feedback is very efficient to regulate the black hole growth,
and radiative feedback can help to balance cooling and modulate the gas dynamics
on the galactic scale.

As is well-known, two ``modes'' of AGN feedback have been identified \citep{Fabian2012, Kormendy2013}. When the mass accretion rate is relatively large, $\ga 0.1 \dot{M}_{\rm Edd}$ (where the Eddington rate is defined as $\dot{M}_{\rm Edd}\equiv 10L_{\rm Edd}/c^2$), we have the radiative or quasar mode. In this regime, a standard thin disk is operating in the central region of AGNs, the luminosity is high, and a strong wind is observed (see \citeauthor{Fabian2012} \citeyear{Fabian2012} for a review). When the accretion rate is lower, the AGN is said to be in the kinetic mode, also known as the ``radio'' or ``maintanence'' mode. This mode corresponds to low-luminosity
AGNs (LLAGNs). In this regime, the accretion flow is expected to be in the hot phase (Narayan \& Yi \citeyear{Narayan1994}, \citeyear{Narayan-Yi1995}; see Yuan \& Narayan \citeyear{Yuan-Narayan2014} for a recent review of the theoretical aspects of this model and its various astrophysical applications). The spectrum from a hot accretion flow is quite different from that of a standard thin disk, as we will describe in more detail in \S\ref{sec:radio-mode-feedback}. Moreover, recent theoretical studies indicate that, in addition to jet, winds also exist in  hot accretion flows \citep{Yuan2012b,Narayan2012,Sadowski2013,Li2013}. Observationally, such a theoretical prediction has been confirmed by the Chandra observation to the accretion flow around the supermassive black hole in our Galactic center \citep{Wang2013}.

The importance of radiative feedback in the quasar mode is obvious and well recognized. However, in the radio mode, the role of radiation is usually assumed to be minor compared to mechanical feedback by winds or jet. The reasons are possibly twofold. First, it is usually thought that the radiative power is low compared to the kinetic power of jet. Second, the efficiency of radiative heating of the emitted radiation on the interstellar medium is assumed to be not high. However, the radiative luminosity may actually be larger than the kinetic outflow power for luminosities $\ga 10^{-4}L_{\rm Edd}$ \citep{Fender2003}. It should be noted that, regarding the radiative heating efficiency, it depends not only on the bolometric luminosity, but also the spectral energy distribution. Taking the Compton heating as an example, the Compton heating is proportional to both
the total radiative flux and the Compton temperature $T_C$ \citep[defined as the effective temperature of the radiation field,][]{Sazonov2005}, and the value of $T_C$ is determined by the spectral energy distribution (see eq. \ref{eq:compton-temperature}): in the quasar mode, typically $T_C\simeq 10^7$K \citep{Sazonov2005}, while in the radio mode,  as we will describe in detail in \S\ref{sec:radio-mode-feedback}, $T_C$ is likely to be as high as $10^9$K, which is about two orders of magnitude higher than that in the quasar mode \citep{YXO2009}. Such a difference of $T_C$ in the two feedback modes has not been considered in the works mentioned above; rather, it is often assumed to maintain the same $T_C$ at different feedback modes.  Therefore, the radiative feedback in the radio mode could potentially be much more important than previous thought. For example, \citet{Novak2011} studied two models with different mechanical feedback efficiency $\epsilon_W$. In the first class of models (``Models A''), $\epsilon_W$ is a constant. In another family (``Models B''), the efficiency decreases with the decreasing luminosity. Although there are some initial evidence that the assumption of Model B leads to more realistic results, \citet{Novak2011} find that Model B predicts too much growth of black hole mass, and incorrect AGN duty cycle. Will the results be changed when we correctly consider a stronger Compton heating in the radio mode? One of our main aims of the present work is to study the effect of a variable $T_C$.

In the present paper, following \citet{Novak2011}, we perform two-dimensional high resolution hydrodynamical numerical simulations to study the effect of AGN feedback on the evolution of an isolated elliptical galaxy. Both mechanical and radiative interaction are considered. Special attention is paid to the effect of radiative feedback. Especially, the Compton temperature is self-consistently determined from the different accretion modes and a high (consistent) $T_C$ is adopted when the accretion is in the hot mode. The paper is organized as follows. In \S2, we introduce the physics of our model, including how do we treat mechanical and radiative feedback, the galaxy model we adopt, and the numerical treatments of stellar mass loss, supernovae, and star formation. The setup and boundary conditions of numerical simulation are described in \S3. Our results are presented in \S4, while \S5 is devoted to summary and discussion.

\section{Model} \label{sec:methods}

In this section we introduce the physics of our feedback model. Many aspects of the simulations such as the galaxy model, the treatment of stellar evolution, and the hydrodynamical equations are the same as in \citet{Novak2011}; for completeness, we briefly recall them.

\subsection{Physics of radiative and mechanical AGN Feedbacks}

\subsubsection{AGN accretion and radiation}

In order to resolve the Bondi radius, we carefully choose the inner boundary of our computational domain. Yet, since the dynamics of accretion within the Bondi radius can't be resolved, we still need to use a sub-grid model to describe the black-hole accretion onto the central AGN. In order to guarantee accretion, in absence of explicit viscosity terms, following \citet{Novak2011}, the specific angular momentum of the gas is assumed to be low. In this case, it is generally assumed that a small disk will be formed at a small circularization radius\footnote{For hot accretion, this scenario is shown to be problematic (Bu et al. 2014, in preparation). But since this does not affect our current study, we still assume this scenario in the present paper.} inside the Bondi radius.

One important aspect to consider is the timescale for the gas at the inner boundary to fall onto the black hole. This is the sum of two timescales \citep{Ciotti2007, Novak2011}: (1) the free-fall timescale $\tau_i$ from the inner boundary to the circularization radius and (2) the accreting timescale $\tau_d$ at the circularization radius, which is determined by the viscous timescale of the small accretion disk. Once the accretion timescale is known, it is possible to evaluate the mass accretion rate onto the black hole. First, we can calculate easily the mass inflow rate at the inner boundary of the simulation domain. As we described in \S1, only a part of the gas can finally reach the black hole because of the mass outflow from the unresolved inner disk. We then need to subtract the
mass outflow rate to get the net mass accretion rate $\dot{M}_{BH}$ at the black
hole horizon and subsequently trace the black hole growth by updating the black hole mass at each time step.

The instantaneous radiative luminosity of the AGN is determined by
\begin{equation}  \label{eq:lum-bh}
L_{\rm BH} = \epsilon_{\rm EM} \dot{M}_{\rm BH} c^2  \, ,
\end{equation}
where $\epsilon_{\rm EM}$ is the radiative efficiency and $c$ is the speed of light. If the accretion flow is described by a standard thin disk (i.e, the AGN is in the quasar mode), the efficiency is only a function of black hole spin and is independent of accretion rate $\dot{M}_{\rm BH}$. We adopt $\epsilon_{\rm EM}\simeq 0.1$ in the present work. When the accretion is in the hot accretion mode (i.e., the AGN is in the radio mode), the value of $\epsilon_{\rm EM}$ becomes smaller and is a function of $\dot{M}_{\rm BH}$. \citet{Xie-Yuan2012} present the detailed study to this problem, and provide fitting formulae of $\epsilon_{\rm EM}$ as a function of $\dot{M}_{\rm BH}$. However, in the present paper, in order to isolate the effect of a time-dependent Compton temperature $T_C$, we simply follow previous works and adopt the following recipe
\begin{equation}  \label{eq:eps-em}
\epsilon_{\rm EM} = \frac{\epsilon_0 A \dot{m}}{ 1 + A \dot{m}},
\end{equation}
with $A=100$ and $\epsilon_0=0.1$ \citep[\citeauthor{Novak2011} \citeyear{Novak2011}, see also][]{Yu2002}. We note that the efficiency described by the above equation is in general lower than the more exact results presented in \citet{Xie-Yuan2012}. In the equation above, the dimensionless mass accretion rate is defined as
\begin{equation}
\dot{m} \equiv \frac{\dot{M}_{\rm BH}}{\dot{M}_{\rm Edd}} = \frac{\epsilon_0 \dot{M}_{\rm BH} c^2}{L_{\rm Edd}},
\label{eq:mdot}
\end{equation}
where $L_{\rm Edd}$ is the Eddington luminosity.

\subsubsection{Radiative Feedback in Kinetic Mode} \label{sec:radio-mode-feedback}

The interaction between the radiation and the galactic ISM proceeds through two channels. One is via radiation pressure, and the other is the radiative heating.  When evaluating radiative heating via Compton scattering, an important concept is the Compton temperature, which is the energy-weighted average photon energy of the radiation emitted by the AGN. It is defined as \citep[e.g.,][]{Park2007}
\begin{equation} \label{eq:compton-temperature}
T_C =\frac{1}{k} \cdot \frac{\int{F_{\nu} h\nu d\nu}}{4 \int{F_{\nu} d\nu}},
\end{equation}
where $F_{\nu}$ is radiative flux, and $k$ and $h$ are the Plank and Boltzmann constants, respectively.
Obviously, the value of $T_C$ is determined by the spectrum of the radiation from the AGN. Sazonov et al. (\citeyear{Sazonov2004, Sazonov2005}) assessed a full range of observational data of quasars, computed the associated Spectral Energy Distribution (SED), and finally calculated the typical Compton temperature of bright quasar, obtaining $T_C\simeq 2.5\times10^7$ K.  Such a temperature is well above the central temperature of the cooling flow gas. So it is natural to expect that radiative feedback from quasars could have profound affects on their host galaxies.
In fact, the radiative heating can be written as
\citep[e.g.,][]{Sazonov2005}
\begin{equation}
H_{\rm Compton}= 4.1\times 10^{-35} n^2 \xi (T_C-T) ~~{\rm ergs}~{\rm cm^{-3}~s^{-1}},
\label{Comptonheating}
\end{equation}
where $\xi=4 \pi F/n$ is the ionization parameter, $F$ is the flux of photoionizing photons and $n$ is the number density of the ISM.

Now the question is what is the value of $T_C$ of the radiation spectrum produced by LLAGNs, i.e., in the radio mode. The previous works adopted the same value as in the quasar mode. However, observations show that the spectra produced by LLAGNs and quasar are quite different, with the most characteristic one being the lack of the ``big blue bump'' in LLAGNs \citep[e.g.,][see also the review by Yuan \& Narayan \citeyear{Yuan-Narayan2014}]{Ho1999, Ho2008, Chiaberge2006, Eracleous2010, Younes2010}. In addition, the spectrum of LLAGNs is more ``X-ray loud'' than that of quasar \citep[refer to Fig. 7 in][]{Ho2008}. This means that the spectra of LLAGNs are much harder than that of quasars, i.e., there is a larger fraction of high-energy photons. \citet{YXO2009} have made an initial calculation to the emitted spectrum from a hot accretion flow and found that the corresponding Compton temperature $T_C$  could be as high as $\simeq 10^9$ K. We would like to emphasize that there are two caveats in this calculation. One is that such value of $T_C$ is calculated from the theoretical spectrum emitted from hot accretion flows. In reality, since the accretion model of LLAGNs also include a truncated thin disk \citep{Yuan-Narayan2014}, we should also include the contribution of such a thin disk. So that the best way is to directly calculate $T_C$ from the observed spectra of LLAGNs at various luminosities. Taking an LLAGN -- NGC3998 --- as an example, we have calculated the corresponding $T_C$ following eq. (4). The multiwaveband spectrum of this source is combined in \citet{Yu2011}. The result is $T_C\approx 10^9$K, although there are some uncertainties in the value of the high-energy cutoff in its X-ray spectrum. So this seems to indicate that the ``theoretical spectrum'' adopted in \citet{YXO2009} is not a bad approximation at least for some  sources, although it is still necessary to check the other sources with various luminosities. The second caveat, which is perhaps more important, is that the relativistic effects are not included in \citet{YXO2009} and eq. (4). Obviously, combining the SED of various AGNs with various luminosities and calculating their  $T_C$ after take into account all relativistic effects will be an important work (Yuan et al., in preparation). Including relativistic effects is likely reduce the value of $T_C$, which will then weaken the radiative heating in the kinetic mode. However, this effect may be cancelled in some degree since the radiative efficiency we adopt in the present paper (i.e., eq. 2) is lower than the exact value calculated in \citet{Xie-Yuan2012}.  Given these complications, in the present work, we simply adopt $T_C \simeq 10^9$ K.  Such a large value of $T_C$ means that even though the bolometric luminosity of LLAGNs is low, its radiative feedback effects are potentially important. This is further strengthened by the possibility that galaxies may spend most of their time in the LLAGN phase.

In this paper we intend to evaluate quantitatively effects of radiative feedback in the radio mode. To this aim we study two types of models, namely model ``f'' and model ``v''. Model ``f'' is for comparison purpose. In this model, the Compton temperature is fixed to be $T_C\simeq 2\times 10^7$K, irrespective of the luminosity of the accretion flow, just as in previous works \citep[e.g.,][]{Ciotti2007, Ciotti2010, Novak2011}.  In model ``v'', the Compton temperature of the spectrum is variable, and is a function of the luminosity of the accretion flow. Specifically, when the accretion rate is high, i.e., the system is in the quasar mode, we adopt $T_C\simeq 2\times 10^7$K; when the accretion rate is low, i.e., the system is in the radio mode, we adopt $T_C\simeq 10^9$K.

The key question is how to treat the transition between the radio and quasar modes. Since the physics of black hole accretion is presumably independent of the black hole mass, we can gain some insight from the observations of black hole X-ray binaries. With the change of accretion rates, this kind of sources transit between two distinct states, namely soft and hard ones. The spectra of the two states are quite different, with the hard state being much harder than the soft state. The accretion flow in these two states are the standard thin disk and hot accretion flows respectively \citep{Done2007,Yuan-Narayan2014}, and the physics of the transition between these two accretion modes is reviewed in \citet{Yuan-Narayan2014}. The dividing luminosity between the two states is $\simeq 2\%L_{\rm Edd}$ \citep{Kalemci2013}, so that in model ``v'' we adopt the Compton temperature as
\begin{equation}\label{eq:model-v}
T_C = \cases{
2.5\times10^7~~ {\rm K},     \quad\quad\quad\quad\,  L/L_{\rm Edd}  >   0.02; \cr
1.0\times10^9~~ {\rm K},     \quad\quad\quad\quad\,  L/L_{\rm Edd} \leq 0.02.    }
\label{twotemperature}\end{equation}

\subsubsection{Mechanical Feedback by Nuclear Winds} \label{sec:wind-model}

In addition to radiation, the nuclear wind from accretion flow is another important channel of AGN feedback. We neglect the role of jet since (at least for an isolated galaxy) it is believed that jets simply drill through the surrounding gas and have little effect to the galaxy. Winds provide mass, energy and momentum rushing into the ISM. The existence of wind in the quasar mode has been directly confirmed by observations, e.g., via the broad absorption lines in some quasars \citep[e.g.,][]{Crenshaw2003,Arav2005}. In the case of radio mode, outflows have also been detected recently in LLAGNs. \citet{Crenshaw2012} investigated the outflow from a sample of nearby AGNs. Of the ten nearby Seyfert 1 galaxies in their sample, six sources still have their bolometric luminosities
below $5\%L_{\rm Edd}$. Their detailed study of the UV and X-ray absorbers clearly shows that a strong outflow exists in these sources. The bolometric luminosity of one source, NGC 4395, is even as low as $10^{-3}L_{\rm Edd}$. Most recently, the existence of outflow in Sgr A* has been confirmed by Chandra observations \citep{Wang2013}. Theoretically, the existence of winds have been confirmed by MHD numerical simulation, although agreement has not been achieved on  the mass flux of winds \citep{Yuan2012b,Narayan2012}.

Following previous works, we adopt a phenomenological approach and notate the properties of wind, namely the fluxes of mass, energy, and momentum, as follows \citep{Ciotti2007,Ostriker2010,Novak2011},
\begin{eqnarray}
\label{eq:mdot-w}  \dot{M}_W &=& 2 \epsilon_W \dot{M}_{\rm BH} c^2  / v_W^2 \, ,   \\
\label{eq:edot-w}  \dot{E}_W &=&\  \epsilon_W \dot{M}_{\rm BH} c^2          \, ,   \\
\label{eq:pdot-w}  \dot{P}_W &=& 2 \epsilon_W \dot{M}_{\rm BH} c^2  / v_W   \, ,
\end{eqnarray}
where $v_W$ is the velocity of the wind, and $\epsilon_W$ is the mechanical efficiency, which
describes the ratio of the wind power and the accretion power. As discussed in detail in
\citet{Ostriker2010}, these expressions guarantee that the mass, energy, and momentum carried by the wind are self-consistent.

The velocity of the wind in the case of quasar mode is relatively well observed, which is roughly $10^4~{\rm km~s^{-1}}$ \citep[e.g.,][]{Crenshaw2003, Chartas2003, Blustin2007, Hamann2008}. In the radio mode, however $v_W$ is poorly constrained by observations. From the theoretical point of view, Yuan et al. \citep[\citeyear{Yuan2012b}; see also][]{Li2013} discuss in a preliminary essay the terminal velocity of wind based on their MHD numerical simulation of hot accretion flow. More systematic study is ongoing (Yuan et al. 2014 in preparation). In the present work, we tentatively adopt the same $v_W$ as
in the quasar mode. As for the mechanical efficiency $\epsilon_W$, even in the case of relatively better observed quasar mode, its value is poorly constrained.

Given this situation, \citet{Ciotti2009b} considered two prescriptions of mechanical feedback, and the corresponding families of models were indicated as ``Models $A$'' and ``Models $B$''.  For models $A$, both the mechanical efficiency (denoted as $\epsilon_W^M$) and the wind opening angles are independent of the accretion rate. In  models $B$, these two quantities vary with the accretion rate, and are arranged to be small at small accretion rates and reach a specified maximum at the
Eddington rate \citep[see also][]{Novak2011}. That is, we have
\begin{equation} \label{eq:eps-wind}
\epsilon_W = \cases{\epsilon_W^M,      \quad\quad\quad\quad\quad\quad\quad\quad\,\,       {\rm [A]}
\cr \displaystyle{{3\over 4}\cdot{\epsilon_W^M ~ l\over 1+0.25 l}},\quad\quad\quad\quad\, {\rm [B]}}
\end{equation}
where $l\equiv L_{\rm BH}/L_{\rm Edd}$. It follows that the efficiency of models
B is in general much lower than in models A. When $l=2$, the mechanical efficiency $\epsilon_W$  of models B is identical to that of models A, but will drop rapidly with the decrease of $l$.

A natural question is then, which one is more realistic among model A and B. In the case of hot accretion flow, in which the outflow is likely mainly produced by MHD mechanism \citep{Yuan2012b}, it is still an open question how the efficiency changes with luminosity. But in the case of radiative line-driven winds, numerical simulations (focused on the inner few hundred $R_g$, where $R_g=G M_{BH}/c^2$ and $G$ is the gravitational constant) indicate that $\epsilon_W$ does fall off when the luminosity decreases \citep{Kurosawa-Proga2009a,Kurosawa-Proga2009b,Kurosawa2009}. In this sense, model B seems to be more realistic.

Following \citet{Novak2011}, we parameterize the angular distribution of the nuclear wind properties as
\begin{equation} \label{eq:ang-distribution}
f_q (\theta) \equiv \frac{1}{Q}\frac{dq}{d\Omega} \,,
\end{equation}
where $q$ is a conserved quantity (mass, energy, or radial momentum), $Q$ is the total amount of the conserved quantity to be injected, and $d\Omega=\sin\theta d\theta d\phi$ is the solid angle covered by winds. We set $f_q (\theta) \propto \cos^2(\theta)$ so that the half-opening angle enclosing half of the mechanical energy is $\simeq 45^{\rm o}$. In terms of solid angle, this means that the wind is visible from $\simeq 1/4$ of the available viewing angles \citep{Proga2004, Kurosawa-Proga2009a}. This fraction is also in agreement with observations of the fraction of obscured and unobscured AGNs \citep{Dai2008, Gibson2009} under the assumption that the two populations are made up of a single population of objects that differ only in viewing angle.

Under the above assumptions, we can conveniently introduce in the numerical simulations the wind feedback by injecting the desired mass, energy, and momentum into the innermost cells of the simulation domain and self-consistently compute the radial transport of these quantities.

\subsection{Galaxy Model: Gravity and Stellar Evolution}
\label{galaxy}

The gravitational potential of the galaxy is contributed to by a dark matter halo and a stellar spheroid embedded in it, with a central black hole. As common, we ignore the self-gravity of the ISM in the simulations. We note that the stellar population is of critical importance in our galactic scale simulations because: (i) The gravity due to the stellar spheroid is dominant at the scale of $0.1-10$ kpc from the galaxy center;
(ii) Abundant gas will be released during the stellar evolution, which is the main gas resource of material fueling the black hole in our simulation. In fact, over a cosmological time span, the evolving stellar population will inject  gas of mass summing up to $\simeq20-30\%$ of the total initial stellar mass into the galaxy, and this injected mass is therefore $\simeq$ two orders of magnitude larger than the black hole mass observed in elliptical galaxies \citep[as $M_{BH} \simeq 10^{-3} M_*$, see][]{Magorrian1998,Pellegrini2012}. The stellar mass losses are thermalized by the relative motion of the stars and ISM \citep[e.g.,][]{Parriott2008}; in addition, supernova explosions provide additional source of mass, metals and energy. A full description of the build-up of the gaseous halo of the galaxy, up to onset of the first ``cooling catastrophe'', has been given elsewhere \citep[e.g.,][]{Ciotti2012}. Here is sufficient to recall that when the ISM density reaches a critical value, suddenly a large amount of cooling gas falls toward the center, a part of it accretes onto the central black hole, a part forms stars, and the remaining is expelled from the central region, until a new cycle repeats. In a word, the stellar population not only determines the source of the accreting gas but also is important to the gaseous dynamics and to the black hole growth.

In the present models, we adopt the Jaffe profile for the stellar component,
\begin{equation} \label{eq:jaffe-dstar}
\rho_*  = {M_* r_*\over 4\pi r^2( r_*+r)^2},
\end{equation}
where $ M_* $ and $ r_*$ are the total stellar mass and the scale-length of the galaxy,
respectively.
Observations show that the total density distributions of ellipticals can usually be well
described by a $r^{-2}$ profile over a large radial range \citep{Rusin2005, Czoske2008, Dye2008}. For this reason, we set the density profile of dark matter halo so that the total mass profile decreases as $r^{-2}$. Finally, the initial black hole mass is determined according to the Magorrian relation \citep{Magorrian1998}.
The velocity dispersion field of the stellar population is characterized by a central projected velocity dispersion $\sigma_o$, and all the other dynamical quantities relevant for the simulations are given elsewhere \citep[e.g.,][]{Ciotti2009a}.

We calculate the stellar evolution \citep[according to stellar evolution theory, e.g.,][]
{Maraston2005} during the simulations following \citet{Ciotti2007, Ciotti2012} \& \citet{Ostriker2010}. In practice, the mass loss rate from the evolving stellar population declines as $\dot{M}_\ast \propto t^{-1.4}$, and it is injected into the computational domain proportionally to $\rho_\ast(r)$. Stellar evolution contributes further to the gas dynamics via type Ia supernova (SN Ia) explosions, which are the results of degenerate white dwarfs, e.g., accreting white dwarfs which finally reach the
Chandrasekhar limit and mergers of two white dwarfs \citep{Ciotti1991}. We assume that each SN Ia ejects $10^{51}$ erg of energy and $1.4~\Msun$ of material into the ISM. As anticipated, the SN Ia rate evolves secularly as $t^{-1}$, so that the specific heating of the injected gas increases with time.

The metal rich ISM is an ideal place for the onset of radiative cooling instability and this, in combination with AGN feedback, leads to inevitable recurrent starbursts \citep{Ciotti2007}. In the simulations, we compute the star formation rate at each radius $r$ using the standard Schmidt-Kennicut scheme. Among the newly formed stars, there is a  population of massive stars whose masses are $> 8 \Msun$.  The massive stars have a relatively short lifetime and will finally evolve to type II supernovae (SNe II) on a timescale of $\simeq 2\times10^7$ years. As for SN Ia, we assume each SN II ejects $10^{51}$ erg of energy into the ISM and leaves behind a neutron star of $1.4~\Msun$.

\subsection{Hydrodynamics}

In our simulations, the evolution of the galaxy under the effect of the central AGN is described by the following time--dependent Eulerian equations of hydrodynamics (see \citeauthor{Ciotti2012} \citeyear{Ciotti2012} for a full description):
\begin{equation} \label{eq:massconsvr}
   \frac{\partial \rho}{\partial t} + \nabla\cdot(\rho{\bf v})
        = \alpha\rho_* + \dot{\rho}_{II} - \dot{\rho}_*^+,
\end{equation}
\begin{equation} \label{eq:momconsvr}
   \frac{\partial {\bf m}}{\partial t} + \nabla\cdot({\bf m v})
        = - \nabla p + \rho {\bf g} -\nabla p_{\rm rad} -\dot{\bf m}^+_* ,
\end{equation}
\begin{equation} \label{eq:engconsvr}
   \frac{\partial E}{\partial t} + \nabla\cdot(E{\bf v})
        =  -p \nabla \cdot {\bf v} + H - C  + \dot{E}_S +\dot{E}_{I}+\dot{E}_{II} -\dot{E}^+_*,
\end{equation}
\noindent where $\rho$, ${\bf m}$ and $E$ are the gas mass, momentum and internal energy per unit volume, respectively, and ${\bf v}$ is the velocity, $p=(\gamma-1) E$ is the gas pressure, and we adopt an adiabatic index $\gamma=5/3$. $H$ and $C$ are the net rates of radiative heating and cooling respectively, ${\bf g}$ is  the gravitational field of the galaxy (i.e., stars, dark matter, plus the time-dependent contribution of the growing central SMBH). For simplicity, we do not take into account effects of the self-gravity of ISM or the gravitational effect of the mass redistribution due to the stellar mass loss and star formation.
The mass source term $\alpha\rho_*$ is the stellar mass loss rate, and $\dot{E}_S$ corresponds to the thermalization of the stellar mass loss due to stellar velocity dispersion. We let $\dot{E}_{I}$ and $\dot{E}_{II}$ be the feedback rates of energy from type I and II supernovae respectively. Here $\rho_{II}$ is the mass return from short-lived type II supernovae and $\dot{\rho}_*^+$, $\dot{\bf m}^+_*$ and $\dot{E}^+_*$ are the sink terms of mass, momentum and energy due to star formation, respectively. We briefly recall the heating processes below, for more details we refer the readers to Ciotti \& Ostriker (e.g., \citeyear{Ciotti2007,Ciotti2012}).

In the energy equation (eq. \ref{eq:engconsvr}), we include 4 semi-analytic heating terms, i.e., the radiative heating $H$, stellar thermalisation $\dot{E}_S$, and supernovae heating $\dot{E}_{I}$ and $\dot{E}_{II}$, which all are expected in real galaxies. However, it is hard to tell which process dominates over the others, because each heating process plays different roles in different regions, at different time.  With stellar thermalisation $\dot{E}_S$, the ISM could be heated up onto a level of the stellar velocity dispersion --- it is not enough to drive outflow. The heating due to Supernovae (i.e., $\dot{E}_{I}$ and $\dot{E}_{II}$) is able to drive galactic-scale outflow but only on the galactic outskirt as the gravity is weak there, while the radiative heating $H$ could affect the ISM in the whole galaxy, especially in the galaxy centre. As a qualitative comparison, the relative importance of the stellar thermalisation $\dot{E}_S$ is always marginal compared to the supernovae heating, except in the innermost region where the stellar velocity dispersion increases because of the gravity of the supermassive black hole. Regarding the supernovae heating items, SNe Ia heating $\dot{E}_I$ is related to old stellar population, which is a secular process. While SNe II heating $\dot{E}_{II}$ depends on the star formation, which is a transient process. Generally speaking, $\dot{E}_{I}$ dominates over $\dot{E}_{II}$, except where starburst occurs. In brief, the radiative heating $H$ and stellar heating processes ($\dot{E}_S$, $\dot{E}_{I}$ and $\dot{E}_{II}$) play important roles in the inner region and the outskirt of the galaxy respectively, of course the relative importance also depends on the AGN luminosity.

Radiative heating and cooling are computed by using the formulae presented in \citet{Sazonov2005}, which describe the net heating/cooling rate per unit volume of a cosmic plasma in photoionization equilibrium. In particular, Compton heating and cooling, bremsstrahlung loss, photoionization, line and recombination cooling, are taken into account.

The total radiation pressure gradient in eq.(\ref{eq:momconsvr}) can be divided into two parts
\begin{equation}
\nabla p_{\rm rad}=(\nabla p_{\rm rad})_{\rm es} + (\nabla p_{\rm rad})_{\rm photo}.
\end{equation}
\noindent Radiation pressure due to {\it electron scattering} is computed as
\begin{equation} \label{eq:prad_es}
(\nabla p_{\rm rad})_{\rm es}=-{\rho\kappa_{\rm es} \over c} {L_{\rm BH}  \over 4\pi r^2},
\end{equation}
where $\kappa_{\rm es}=0.35~{\rm cm^2~g^{-1}}$ is the electron scattering opacity. The radiation pressure due to all of the radiatively heating processes is calculated simply as (see \citeauthor{Novak2011} \citeyear{Novak2011} for details)
\begin{equation} \label{eq:prad_photo}
(\nabla p_{\rm rad})_{\rm photo} = {H \over c}.
\end{equation}

\section{Model Setup}

Following \citet{Ciotti2007}, we choose the galaxy parameters so that the model obeys to the edge-on view of the Fundamental Plane \citep{Djorgovski1987, Dressler1987} and to the Faber-Jackson (\citeyear{Faber1976}) relation, in particular, the total stellar mass, $\dot{M}_\ast = 3\times10^{11}\Msun$, and the galaxy effective radius $R_e=6.9$ kpc, plus the contribution of the dark matter halo corresponding to a central projected stellar velocity dispersion $\sigma_o=260 {\rm km~s^{-1}}$, and to a stellar mass-to-light ratio $M_*/L_{\rm B}=5.8$ in solar units. The initial mass of the black hole is $M_{BH} = 10^{-3}M_* = 3\times10^8\Msun$.
The gas density is initially set to be a very low value so that the gas in the simulations comes almost exclusively from explicit source terms arising from stellar evolution. This is often termed ``secular evolution'' to distinguish it from evolution induced by cosmological effects (such as galaxy merging).

We perform two dimensional hydrodynamic simulations with {\it ZEUS-MP/2} \citep{Hayes2006} in spherical coordinates ($r$, $\theta$, $\phi$), i.e., all quantities are assumed to be axisymmetric. Following \citet{Novak2011}, in the $\theta$ direction the mesh is divided homogeneously into 30 angular cells, while for the radial direction (covering the radial range of 2.5 pc - 250 kpc), we use a logarithmic mesh with 120 bins to reach the needed resolution in the inner computational region. With this choice each cell is 10\% larger than the previous inner cell.
It is most important to resolve the fiducial Bondi radius so that the accretion rate can be robustly estimated. In particular, the inner boundary radius ($r_{in} = 2.5$ pc) is chosen to be within the Compton and Bondi radii when the Compton temperature $T_C\simeq 2.5\times10^7$ K. The Compton radius $r_C$ is where the Compton temperature is equal to the local virial temperature, mainly determined by $M_{BH}$ \citep{Ciotti2007,YXO2009}. However, $r_C$ is not resolved for high $T_C$, as in models ``v'' when the Eddington ratio is low, i.e., $T_C\simeq10^9$ K and then $r_C \simeq 0.06$ pc with a black hole mass of $3\times10^8\Msun$. The typical values of the Bondi radius (for the central gas temperature and density of elliptical galaxies) range between 10 and 100 pc \citep[e.g.,][] {Pellegrini2010}.  The Bondi radius is estimated by using the sound speed $c_{s,in}$ at the inner boundary,
\begin{equation}
r_B = \frac{G M_{\rm BH}}{c_{s,in}^2}.
\end{equation}
Since $r_B$ varies during the simulations, in the code the radial inflowing velocities at the inner boundary are limited to $c_{s,in} (r_B/r_{in})^2$ in case that the Bondi radius is unresolved. This helps to avoid an unphysical large accretion rate. We stress that in the simulations Bondi accretion {\it is not imposed.}

The simulations in our paper are very time-consuming despite of the low numerical resolution. This is because of the large dynamical range. Technically, it is too computationally expensive to implement both large dynamical range and high numerical resolution simultaneously. 
One of us (Gregory Novak) had tested the model in a previous (but similar) paper with different numerical resolutions and with different dynamical ranges \citep{Novak2011}. In his tests with higher resolution, similar results are found. However, when testing the model with a larger inner boundary (i.e., smaller dynamical range), the results changed significantly. Therefore, in this paper we chose to sacrifice the numerical resolution to cover a large dynamical range.

We assume reflecting boundary conditions on the $\theta$ boundaries occurring at each pole.  On the inner/outer radial boundary, we use the standard ``outflow boundary condition'' in the {\it ZEUS} code ( see \citeauthor{Stone1992} \citeyear{Stone1992} for more details). This allows both outflow and inflow depending on the state of gas just outside/inside the inner/outer boundary.

We use a semi-implicit scheme for radiative cooling because the cooling timescales could be very short, e.g., in galactic center where gas density could occasionally be very high.  If the time step required by the Courant-Friedrich-Levy (CFL) condition is shorter than the cooling time, the code updates gas temperature due to radiative cooling explicitly. Otherwise we use an implicit scheme to update the energy integration.
To avoid unphysical cooling and dramatic collapse which usually cause the code to crash, we also impose three limits dictating that the gas not drop below the temperature of the cosmic microwave background, the effective temperature associated with the AGN radiation field (note that this temperature is not equal to $T_C$) or the effective temperature of the stellar radiation field. In addition, we set a temperature floor of $10^4$ K in the cooling functions, since the gas cannot reach these low temperatures by radiative cooling alone \citep{Sazonov2005, Novak2011}.

\section{Results} \label{sec:results}

As discussed in the Introduction, in this paper we study the effects of AGN radiative feedback in both the quasar and radio modes, paying special attention to the effects of a variable $T_C$ in the two modes. For this aim, we have run two groups of simulations according to the behavior of $T_C$, namely models ``f'' and ``v''. In models ``f'' $T_C$ is fixed, while in models ``v'' $T_C$ changes according to eq.(\ref{eq:model-v}). Table 1 shows various models we have run. Capital letters ``A'' and ``B'' in the model names represent the two types of the nuclear
wind model adopted (see eq.\ref{eq:eps-wind}), while the numbers denote different mechanical efficiency $\epsilon_W^M$ of the wind, with increasing numbers corresponding to decreasing values of $\epsilon_W^M$. In the following we first present an overview of the results, then we focus on some more specific property of the models.

\subsection{Overview of results}\label{sec:overview}

\begin{figure*}[!ht]
\begin{center}
\includegraphics[width=0.26\textwidth]{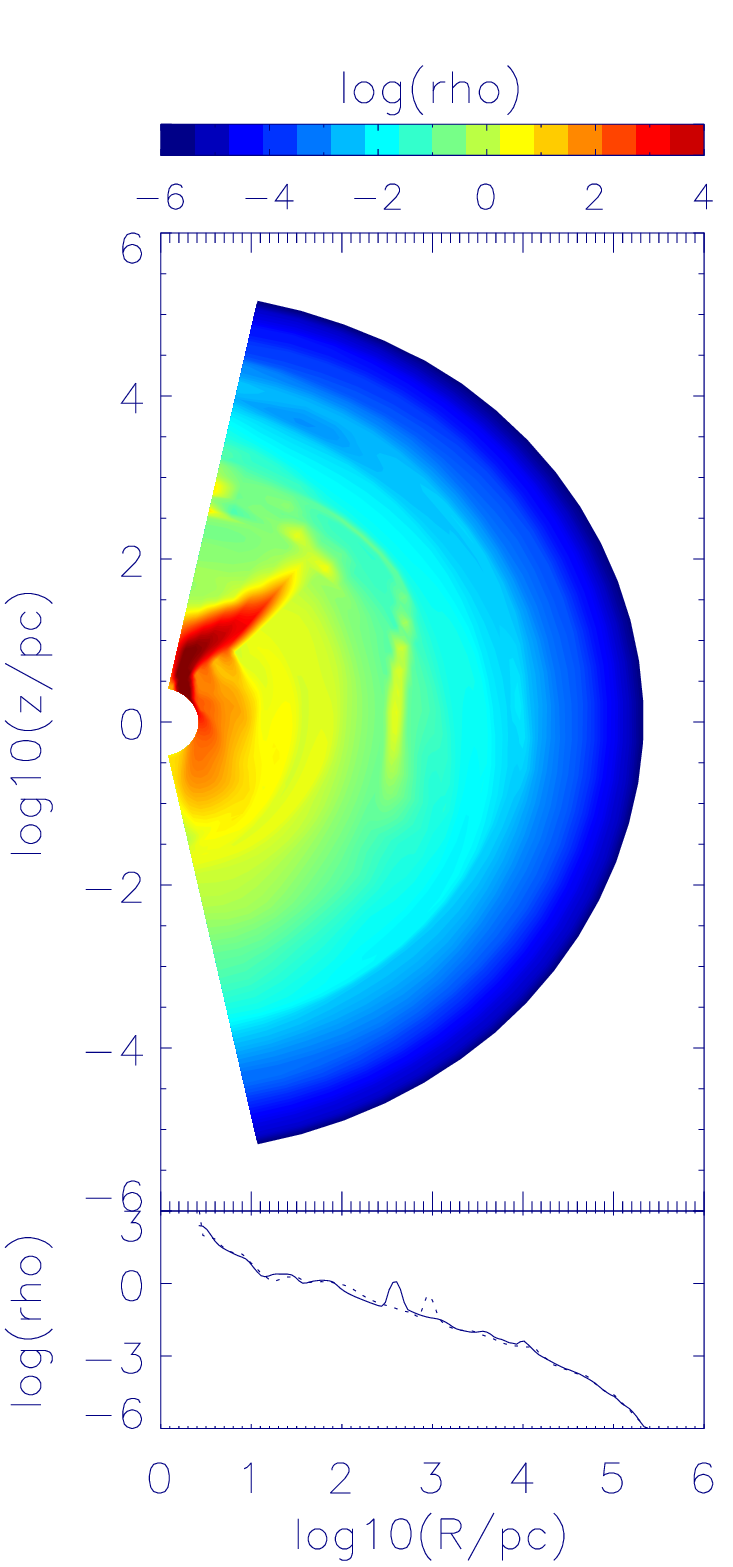}
\includegraphics[width=0.26\textwidth]{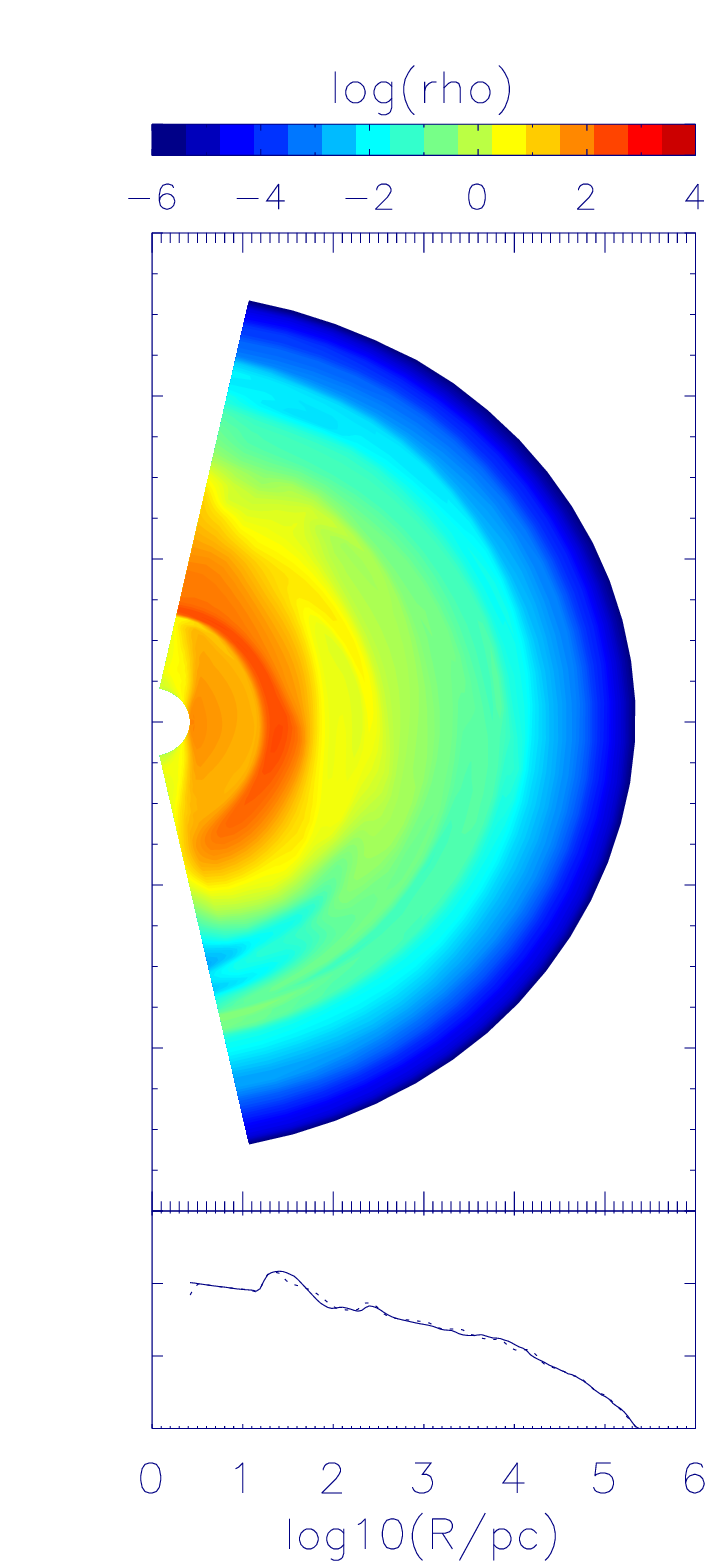}
\includegraphics[width=0.26\textwidth]{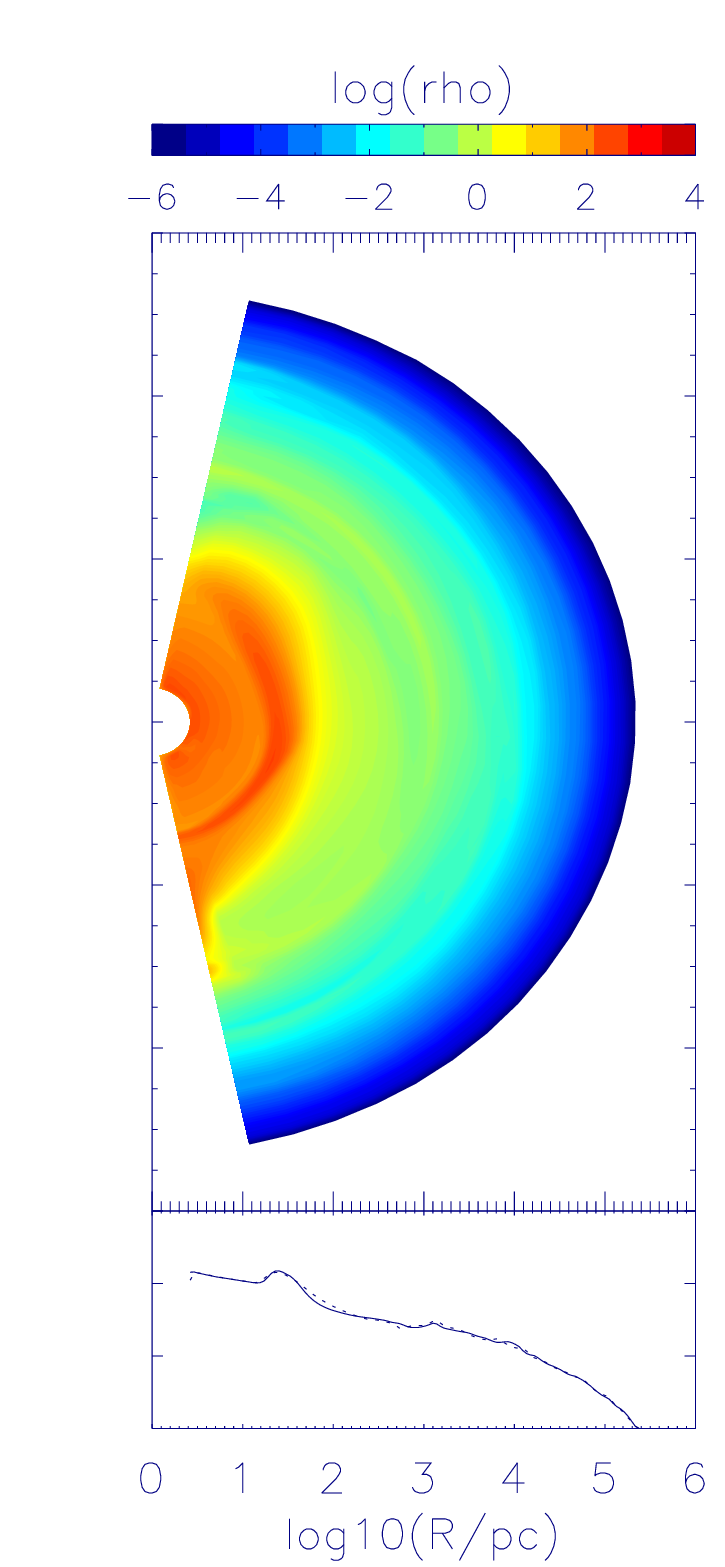}
\includegraphics[width=0.26\textwidth]{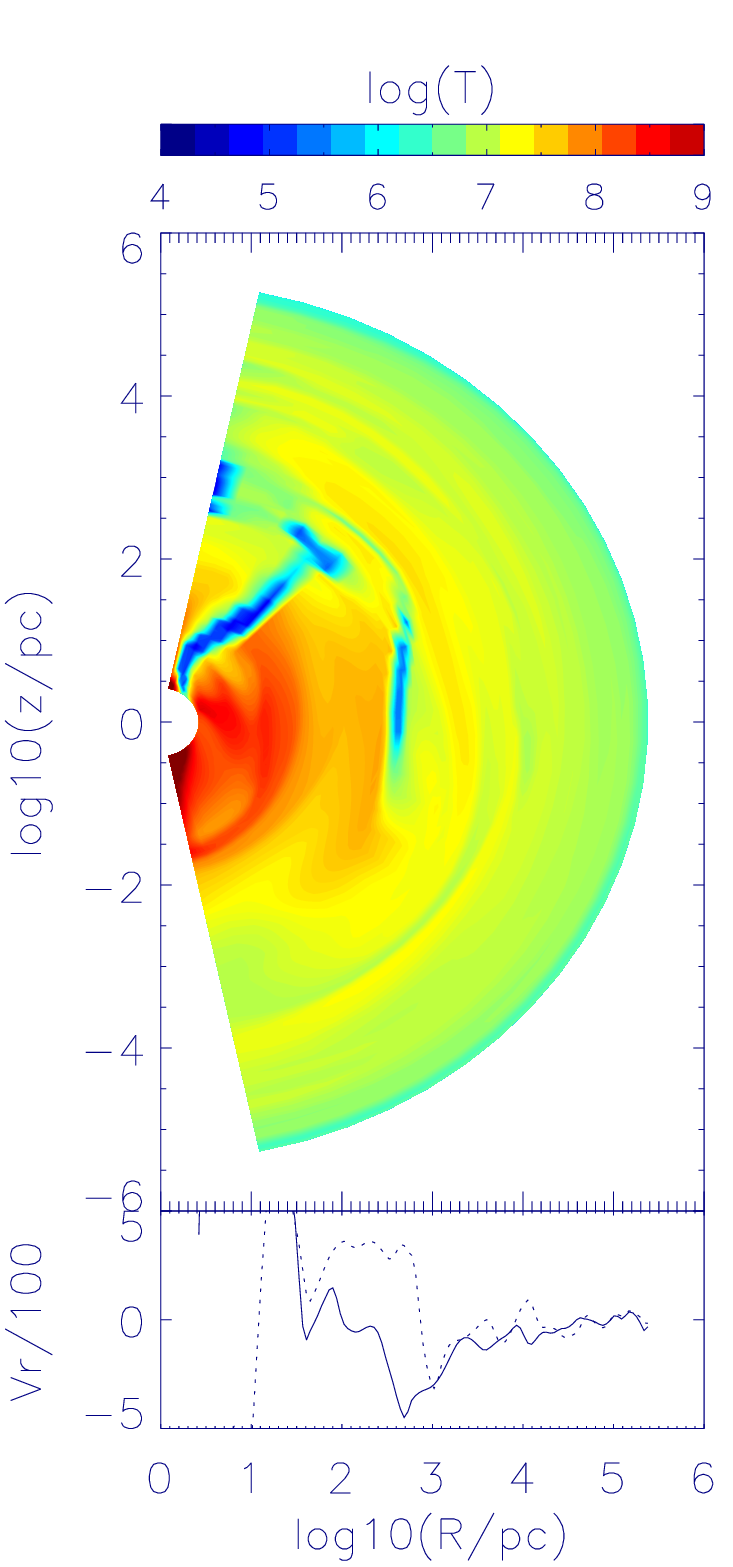}
\includegraphics[width=0.26\textwidth]{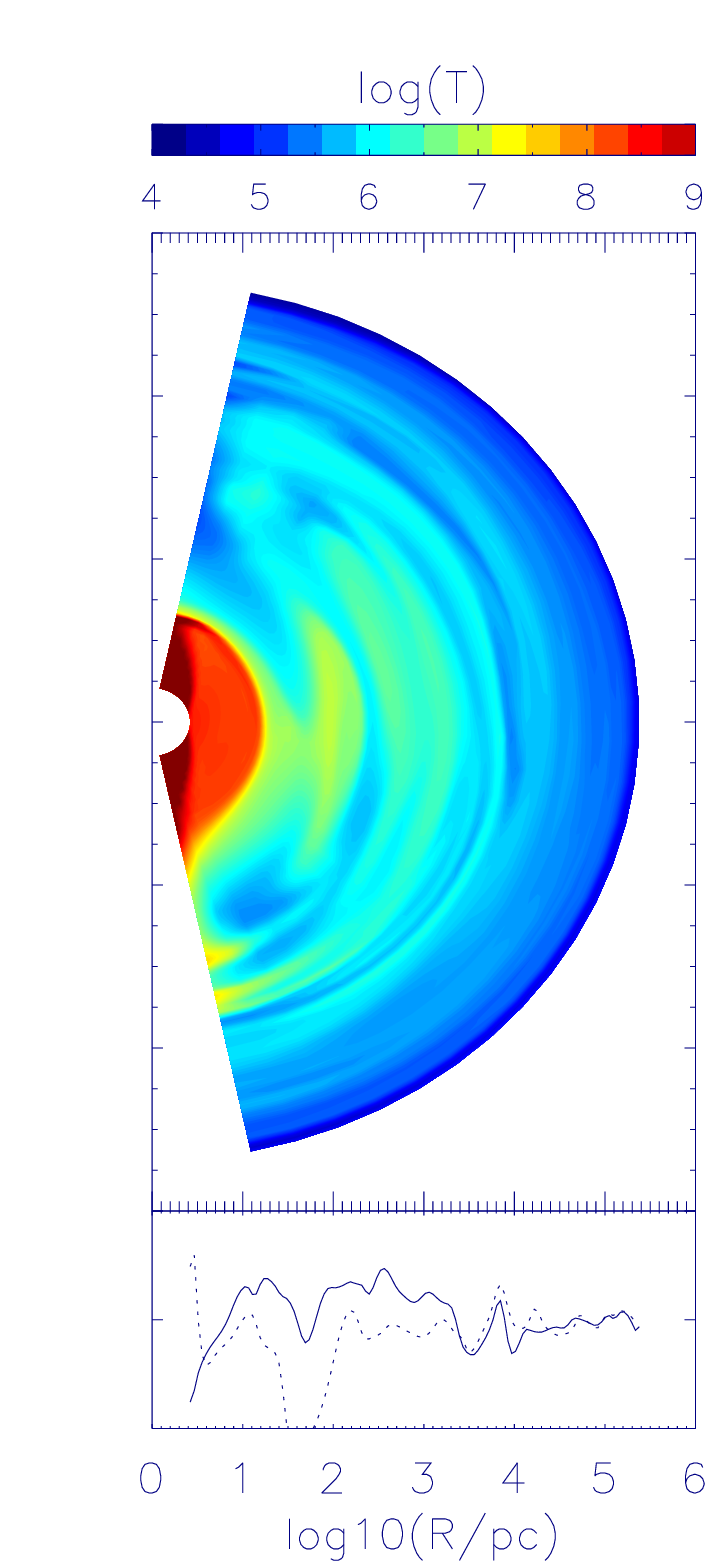}
\includegraphics[width=0.26\textwidth]{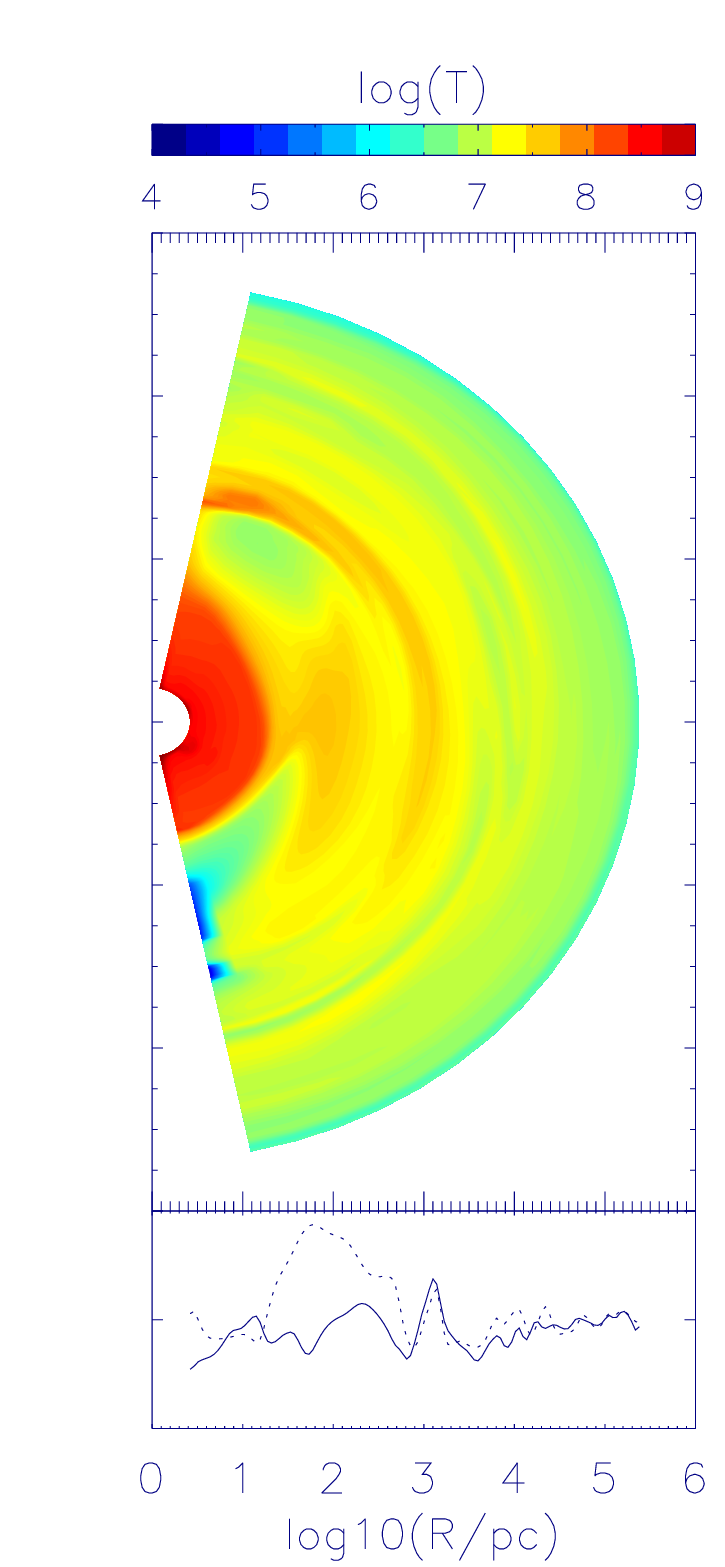}
\end{center}
\caption{Snapshots of the ISM profiles just before ($t=9.3$Gyr, left panel), during ($t=9.3125$Gyr, middle panel) and after ($t=9.325$Gyr, right panel) an AGN burst in model B05v. The upper panel shows the pseudocolor contours of density and the radial profiles of the gas density at two layers of grids just above (solid) and below (dashed) the equatorial plane. The lower panel shows the pseudocolor contours of temperature and the radial profiles of radial velocity at two layers of grids just above (solid) and below (dashed) the equatorial plane. In the contours warmer colors stand for higher values and note that the axises are logarithmic distances.}
\label{fig:snapshot-B05v}
\end{figure*}

In general, the simulation results are similar to those presented in \citet{Novak2011}. For examples, Figure \ref{fig:snapshot-B05v} shows the snapshot of the density and temperature of the interstellar medium at different epochs for model B05v used here as a reference model, being the results of other models similar. At early times, the ISM is very tenuous and the galaxy is in a quiescent phase, where only the passive evolution of stellar population is taking place. Stellar mass losses enrich the gas content of the galaxy gradually. Meanwhile, such gas is heated to a temperature near (slithtly higher) the local virial temperature as determined by the local stellar velocity dispersion and by type Ia supernova explosions. In this evolutionary phase, the accretion onto the black hole is very low, and it is Bondi-like. The black hole remains quiescent until the ISM is globally cooled via radiative cooling. This catastrophic cooling leads to collapse of ISM toward the center, i.e., a ``cooling flow'' occurs, beginning with the formation of a cold shell of $\simeq$ kpc radius \citep{Ciotti2007}. Such a cold shell is unstable due to the Rayleigh-Taylor instability and to be disrupted quickly. On the path towards to the galactic center, the gas becomes even denser both due to the volume effect and to the accumulation of gas by the collapsing cold shell. The increase of gas density in turn enhances radiative cooling, which is proportional to density squared. Such a cool and dense environment is
ideal for star formation or even a starburst (Fig.\ref{fig:snapshot-B05v} left panel). Actually, star formation is a very significant process of consuming the accreting gas (about 1/3 of the total) before it reaches the black hole. Finally the black hole gets the fuel and AGN activity is triggered (Fig.
\ref{fig:snapshot-B05v}, middle panel).

Once the central AGN is accreting at a significant level, it produces radiation and a nuclear wind. They interact with the interstellar medium in the galaxy, which in turn regulates the central AGN feeding. For radiative feedback, the ambient gas could be heated almost immediately because of the irradiation by AGN via, e.g., inverse Compton scattering and photoionization. The mechanical feedback by outflow/wind is important mainly in the region very near to the black hole as mass flow can immediately mix into the ambient ISM, affect the gas dynamics by ram pressure and heat the gas via shocks. As shown by the right panel of Fig. \ref{fig:snapshot-B05v}, when the ISM is either heated or pushed outward as the consequence of  radiative and mechanical feedback, gas is expelled from the galactic center, leaving behind a hot cavity. Finally, a new loop of nucleus activity starts over again when the stellar mass losses replenish the galaxy, a critical density for catastrophic cooling is reached, and a new cold shell forms and collapses. A complete discussion of the life cycle of cold shells, their evolution under the action of radiative losses, AGN feedback, star formation, can be found in \citet{Ciotti2007} and \citet{Ciotti2009b,Ciotti2010}. \citet{Jiang2010} also studied the synchrotron emission produced by the AGN bursts, that behave like a giant supernova remnant. We note that AGN feedback is not only {\it negative}, quenching the cooling flow, but could also be {\it positive} in some situations, inducing star formation in the central region of the galaxy \citep{Ciotti2007,Liu2013b}

\subsection{Light curve of AGN luminosity}

\begin{figure*}[ht]
\begin{center}
\includegraphics[width=0.6\textwidth]{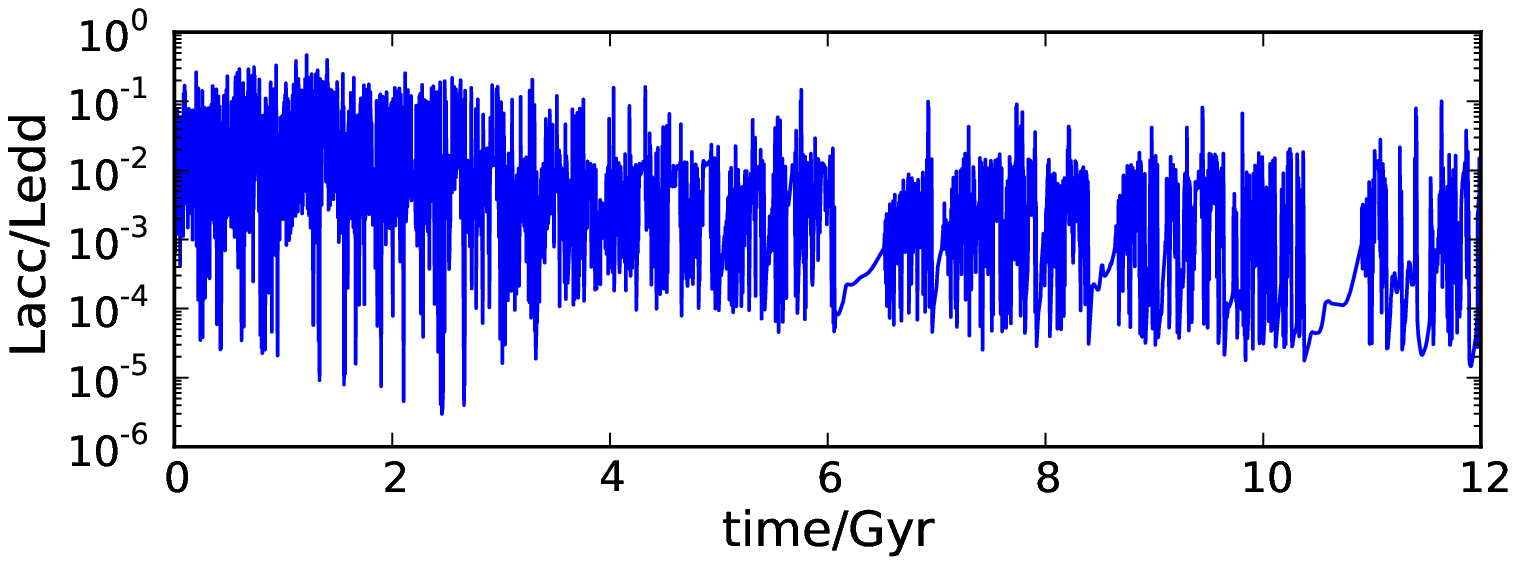}
\includegraphics[width=0.6\textwidth]{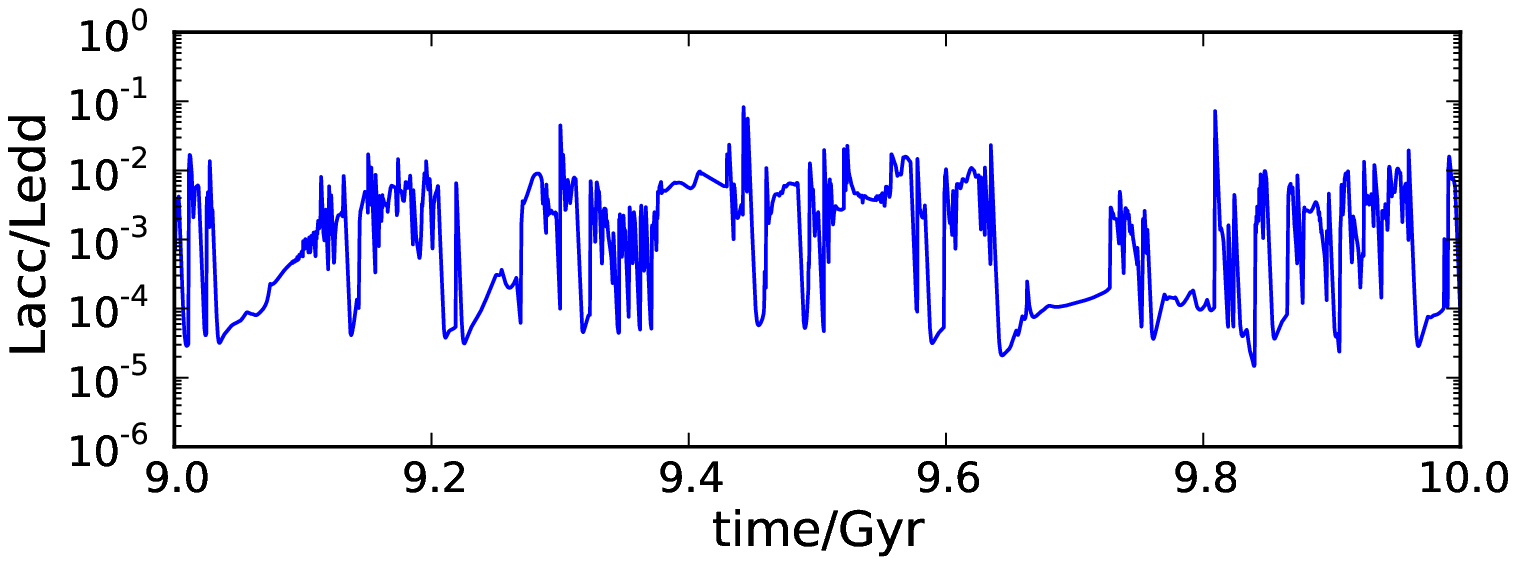}
\includegraphics[width=0.6\textwidth]{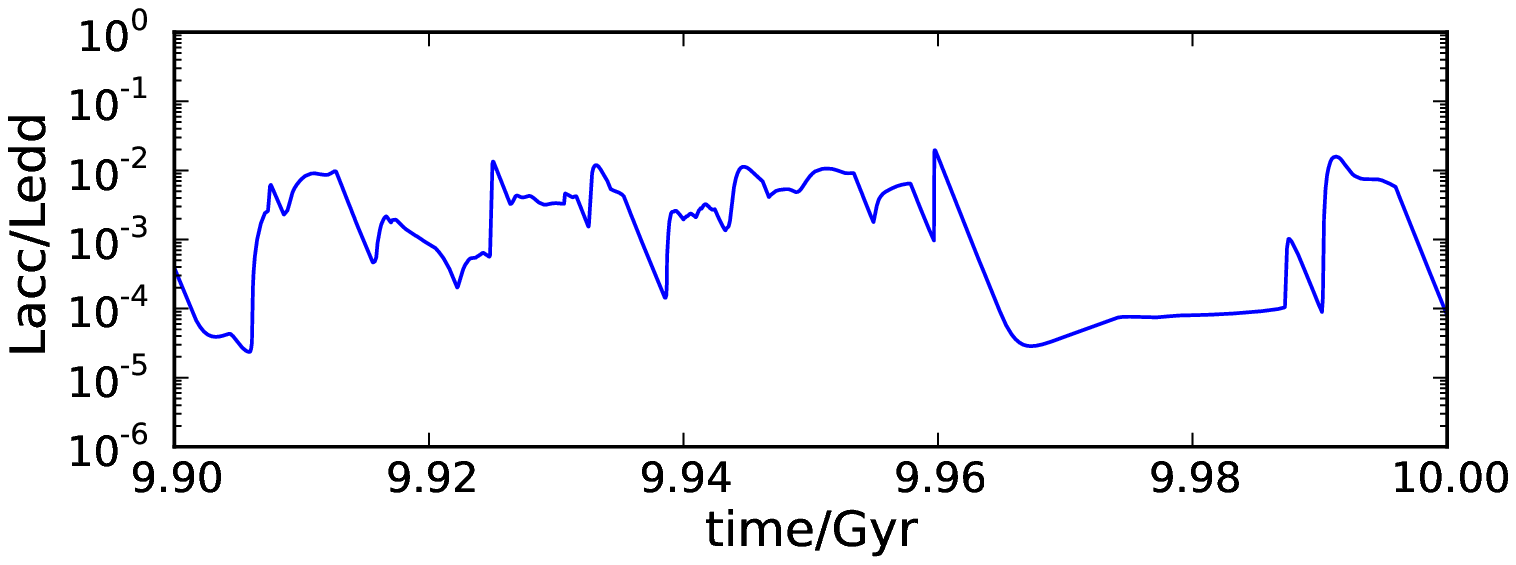}
\end{center}
\caption{The light curve of AGN luminosity during the cosmological evolution in model B05v. The three panels correspond to the results with different time resolution.}
\label{fig:acc-his-b05v}
\end{figure*}

The light curve of the AGN luminosity over the whole cosmic epoch  has been shown in \citet{Novak2011} for Model A. Figure \ref{fig:acc-his-b05v} shows the light curve for a ``B'' type model (in particular model B05v). The three panels correspond to different time resolutions. By comparison with Fig. 6 in \citet{Novak2011} it can be seen that the overall luminosity evolution of models A and B is quite similar. The black hole accretion is very chaotic with a complicate temporal substructure, and many feedback loops are observed. A decline of the overall profile is clearly shown (top panel). Since the luminosity is shown in unit of $L_{\rm Edd}$, such a decline is mainly due to the mass growth of the central black hole. The luminosity evolution reflects the hydrodynamical phases briefly summarized in \S\ref{sec:overview}. We can see that the luminosity valleys are followed by a relatively long-term climb (middle panel). These are dividing points of feedback loops and signatures of strong AGN feedbacks. AGN feedback heats the ISM, expel it out of the galactic center and stop the nuclear activities (Fig. \ref{fig:snapshot-B05v} right panel), until a new cooling cycle begins followed by black hole accretion. In the bottom panel, we can see the luminosity fluctuations on short time scales, i.e., the temporal substructure within a single accretion event. Rayleigh-Taylor instability and the interaction of direct and reflected shock waves \citep[see \citeauthor{Ciotti2007} \citeyear{Ciotti2007} for a full discussion, see also][]{Ciotti2010} are the main causes of such fluctuations. In fact, in Fig. \ref{fig:snapshot-B05v} (left panel) we can see that a cool dense shell forms around the galactic center: the density in the shell is higher than that of its inner gas, while the gravitational force points to the center. As a result, the cold shell fractures and falls onto the center piece by piece asynchronously, which finally induces the quick fluctuations in the light curves.

\begin{figure*}[!ht]
\begin{center}
\includegraphics[width=0.6\textwidth]{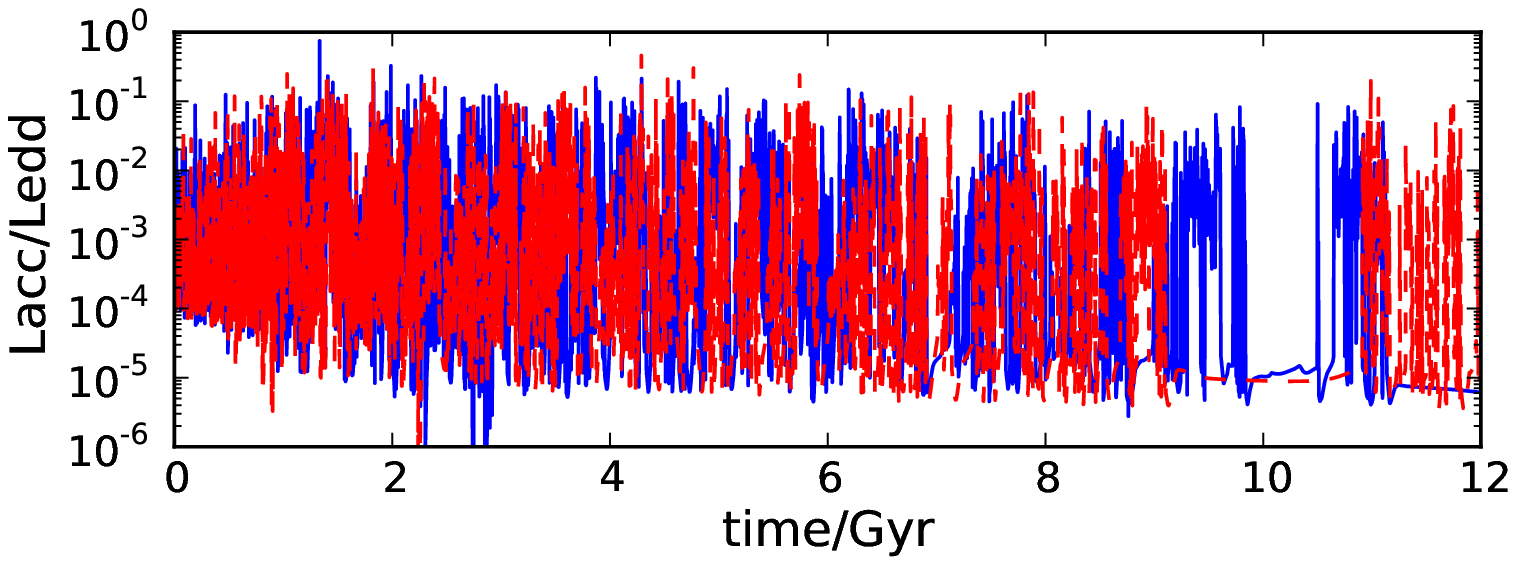}
\includegraphics[width=0.6\textwidth]{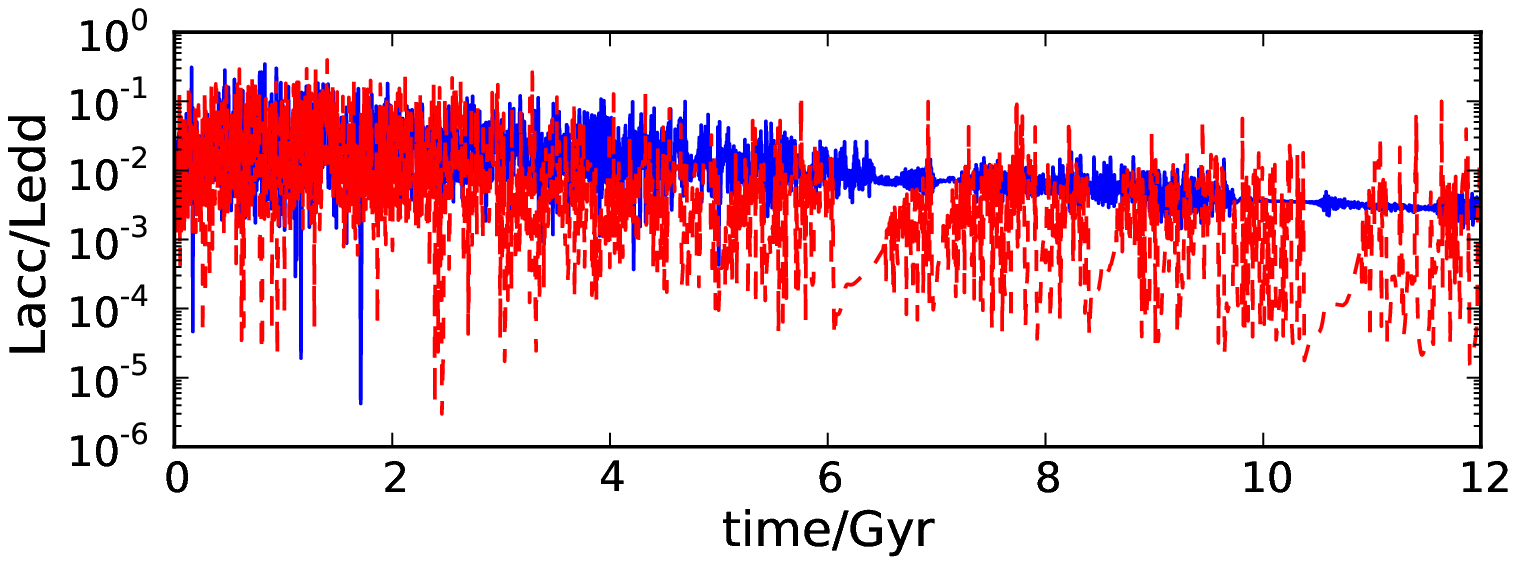}
\end{center}
\caption{Comparison of the light curves of AGNs in different models in the cosmological evolution. {\it Upper panel:} Models A1v (red dashed lines) and A1f (blue solid lines); {\it Lower panel:} Models B05v (red dashed lines) and B05f (blue solid lines). Recall that the red (v) models have enhanced Compton heating and the lower panel (B) has less efficient wind driving at low luminosity which would occur for radiatively driven winds.}
\label{fig:acc-his-a1}
\end{figure*}

Figure \ref{fig:acc-his-a1} compares the different light curves of different models, namely A vs. B and f vs. v. The upper panel shows the light curves of model A1v (red dashed line) and model A1f (blue solid line). Both of the light curves are chaotic, and there is little qualitative difference between the two models. This is not surprising because in models ``A'' the mechanical AGN feedback by nuclear winds is independent of $L_{\rm BH}$, so that the change of radiative feedback (f vs. v) is not important. The behavior of the light curve changes for Model B whose mechanical efficiency is smaller thus mechanical feedback is weaker.  It can become not chaotic at all, as shown by the blue solid line in the lower panel of Fig. \ref{fig:acc-his-a1} (for model B05f).  However, the light curve shown by the red dashed line (model B05v) is again chaotic. This is because in this model, although the mechanical feedback is weaker, the radiative feedback becomes stronger due to the consideration of a higher $T_C$ in the kinetic mode. In addition, during most epochs of the evolution, the accretion luminosity is significantly lower in model B05v than in model B05f. This result is because of the stronger radiative heating in model ``v'' due to the higher Compton temperature which results in a overall smaller accretion rate during the evolution.

\subsection{Radiative efficiency}

In the third column of Table \ref{tab:Model-Summary}, we list the mass-weighted mean
radiative efficiencies $\left< \epsilon_{EM} \right>$ of the black hole accretion flow
averaged over the evolution time as
\begin{equation}\label{eq:averaged_efficiency}
\left< \epsilon_{EM} \right> \equiv \frac{\int{L_{\rm BH}~dt}}{\int{\dot{M}_{\rm BH}~c^2~dt}}.
\end{equation}
Note that if the bulk of black hole accretion is in quasar mode, then according to eqs.(\ref{eq:eps-em})-(\ref{eq:mdot}) one expects $\left< \epsilon_{EM} \right> \simeq \epsilon_0 = 0.1$. Instead, if low luminosity accretion is important, $\left< \epsilon_{EM} \right> < 0.1$. In all of the models except models A0f and A0v, $\left< \epsilon_{EM} \right>$ lies in a narrow range $\simeq 0.056-0.074$. This efficiency is similar to that found by \citet{Kulier2013} applying the Soltan argument \citep[\citeyear{Soltan1982}, see also][]{Yu2002,Haiman2004} but correcting for orbiting and ejected black holes. The values of $\left< \epsilon_{EM} \right>$ in model A0f and A0v
are the smallest. This is because of the strong mechanical feedback: the mechanical efficiency $\epsilon_W$ in these two models is the largest among all models. Such a strong wind is very efficient in regulating mass accretion, since it pushes the gas away, thus the accretion rates are on average lower compared to other models. When the accretion rate is low, the efficiency will also become low.

In general, the value of $\left< \epsilon_{EM} \right>$ is slightly larger in models ``v'' than in ``f''. The reason is as follows. Refer to eq. \ref{eq:averaged_efficiency} the numerator is obviously dominated by the contribution in the quasar mode, which is roughly same for the models ``v'' and ``f'' because the change of $T_C$ only applies in the low luminosity radio mode. The denominator is precisely equal to the growth of the black hole, i.e, the fourth column in Table \ref{tab:Model-Summary}. Compared to model ``f'', the Compton heating in model ``v'' is stronger due to the higher $T_C$. When the gas temperature is higher, the mass accretion rate will be lower. This results in an overall smaller black hole growth over the cosmological time, as discussed in the following section.

\subsection{Mass growth of the black holes}
\label{blackholegrowth}

The regulation of black hole growth at the centers of spheroids is an important effect of AGN feedback. In fact, we recall that also in isolated galaxies (i.e., no merging), the gas produced by the evolving stars over the whole evolution is more than two orders of magnitude larger than the observed masses of black hole in the local universe. The fourth column in Table \ref{tab:Model-Summary} is the mass growth $\Delta M_{BH}$
of the supermassive black hole during the whole evolution epoch.

For both model ``v'' and ``f'', we can see that the black hole growth is quite sensitive to the mechanical efficiency $\epsilon_W^M$, namely $\Delta M_{BH}$ drops rapidly when $\epsilon_W^M$ increases. Physically, this is because with the increase of $\epsilon_W^M$, the gas surrounding the black hole is pushed away more strongly, thus the mass accretion rate decreases. Interestingly enough, we find that there exists a critical $\epsilon_W^M\simeq 10^{-4}$, below which the black hole growth $\Delta M_{BH}$ is insensitive to $\epsilon_W$. This is the case of  models B3f and B4f (or B3v and B4v). The explanation is that below such a small value of $\epsilon_W^M$, the AGN feedback becomes dominated by radiative feedback.

In the regime of low mechanical efficiency ($\epsilon_W^M<10^{-4}$), we found too much black hole growth. For example, in models B3f and B4f the final mass of the black hole is up to $10^{10}\Msun$, which is beyond the typical mass range of a supermassive black hole. In the regime of high mechanical efficiency ($\epsilon_W^M>10^{-3}$), the black hole growth is heavily suppressed, e.g. in models A0f and A0v, which results in a too low level of AGN activity as argued in \citet{Novak2011}. Thus we arrive at the important conclusion that the model-favored mechanical efficiency, for both type ``f'' and type ``v'' model, lies in a range of $\simeq10^{-3}-10^{-4}$, which is consistent with the previous works and observational suggestions \citep[e.g.,][see also \citeauthor{Gaspari2012}\citeyear{Gaspari2012}; \citeauthor{Tombesi2013} \citeyear{Tombesi2013}]{Ciotti2009b,Ciotti2010,Ostriker2010}. It is interesting to note that such a mechanical efficiency seems to be roughly consistent with the MHD numerical simulation to the outflow in hot accretion flows \citep[e.g.,][eq. 34 therein]{Yuan2012b}.

In the case of fixed Compton temperature, we find that the mass of the black holes in models Bf are all larger than in the corresponding models Af. This is obviously because of the stronger mechanical efficiency $\epsilon_W$ in model A than in model B for given $\epsilon_W^M$. For example, as shown in Table \ref{tab:Model-Summary}, the final black hole mass in A0f is smaller than that in B0f by nearly two orders of magnitude, and is nearly three orders of magnitude smaller than  in model ``nowind.f'', in which there is no nuclear wind at all. From Table \ref{tab:Model-Summary}, we can see that sometimes the final mass of the black hole can be as large as over $10^{10}\Msun$. This obviously violates the observations, and indicates that when a fixed $T_C$ is adopted models B tend to overpredict the final black hole mass, although they may be more realistic than models A as we have mentioned in \S1. This conclusion is consistent with that in \citet{Novak2011}.

It is therefore of great importance that the black hole growth $\Delta M_{\rm BH}$ in models ``v'' with variable Compton temperature is systematically smaller. This is true for both models A and B. This is because a higher $T_C$ is adopted in model ``v'' in the radio mode. A higher $T_C$ implies that the gas surrounding the black hole is heated to a higher temperature, the mass accretion rate is smaller and less material is accreted by the black hole. This conclusion is also reinforced by the fact that when $\epsilon_W$ is smaller, the discrepancy of $\Delta M_{\rm BH}$ between models ``v'' and ``f'' is larger. This is because when $\epsilon_W$ is smaller, the role of mechanical feedback becomes less important compared to the radiative feedback therefore  the effect of $T_C$ will be more significant.

An important result in this context is that even for Model B, when we consider a higher $T_C$ in the hot accretion phase (i.e., kinetic mode), i.e., model ``v'', the final mass of the black hole is not unrealistically large. The effect of a changing $T_C$ is so relevant that it is testified also by the limiting case model ``nowind.v'', in which there is no wind feedback at all and the mass of the black hole is the largest, however also in this case the final mass of the black hole remains below $10^{9.5}\Msun$. Such a range of black hole mass is fully consistent with observations, thus models ``B'' are ``revived''.

\subsection{AGN duty cycle}

\begin{figure*}[ht]
\begin{center}
\includegraphics[width=0.75\textwidth]{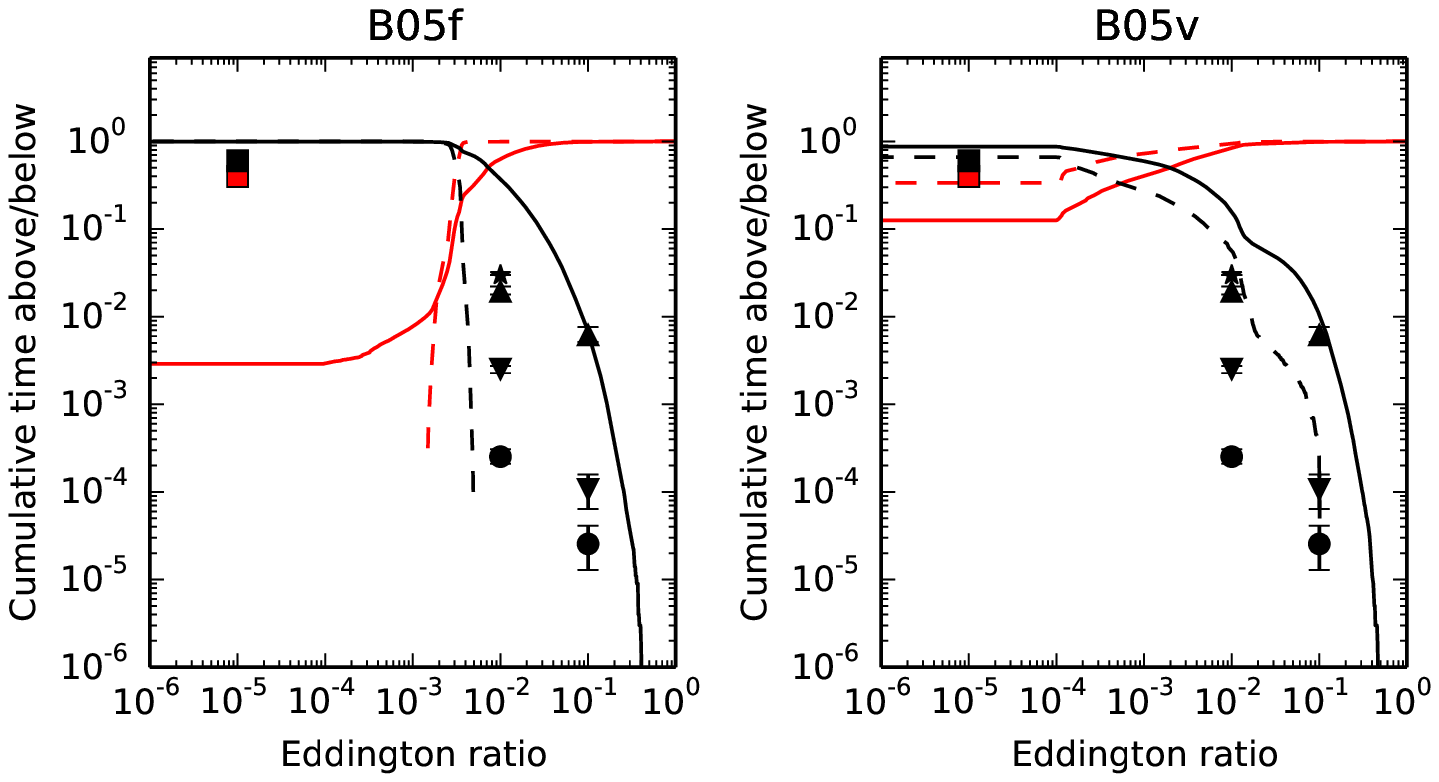}
\includegraphics[width=0.75\textwidth]{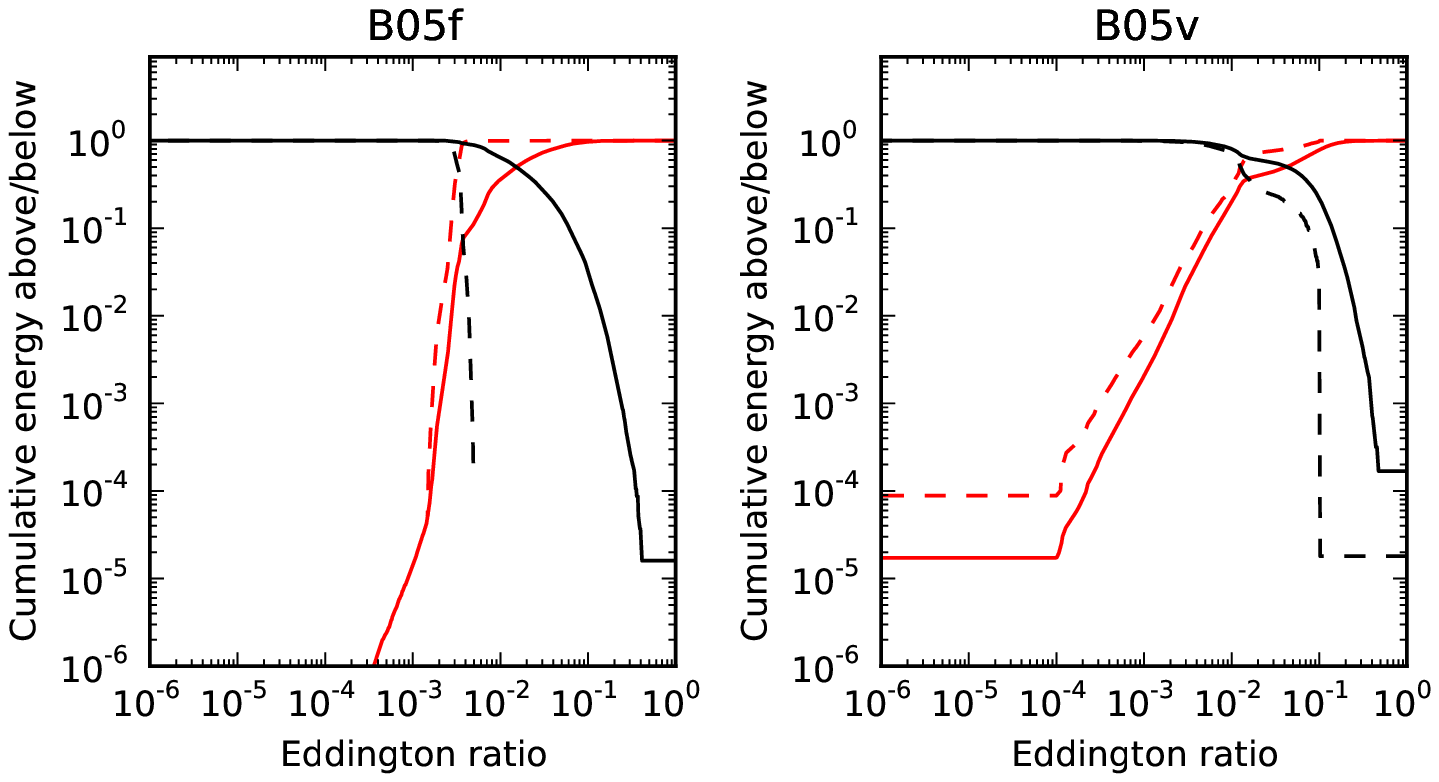}
\end{center}
\caption{The AGN duty cycle as a function of Eddington ratio for models B05f (left panels) and B05v (right panels). In the top panels, the black/red lines show the fractions of cumulative time above/below the given Eddington ratio. Solid lines are the results compiled from the entire simulation time, while dashed lines only use the data in the final 2 Gyr. The points are observational constraints. The black/red points are the measurements of the faction of objects above/below the given Eddington ratio. The squares, cycles, upward- and downward- pointing triangles are from \citet{Ho2009}, \citet{Greene-Ho2007}, \citet{Kauffmann-Heckman2009} and \citet{Heckman2004}, respectively, which are all compiled from low-redshift observations and should be compared with dashed lines. The star is a constraint compiled from high-redshift observations by \citet{Steidel2003}, which should be compared with solid lines. In the bottom panels, the lines are of the same meanings as in the top panels, while they show the fractions of cumulative radiative energy during the evolution.}
\label{fig:duty_cycle-b05}
\end{figure*}

One of the most important results that \citet{Novak2011} found is that for models B, in addition to the problem of black hole growth, also the AGN duty cycle is not consistent with observations. When a high $T_C$ is adopted as in our model ``v'', the dynamics of the gas fueling the black hole and subsequently the mass accretion rate is completely changed. As a consequence, also the AGN duty cycle changes. In Fig. \ref{fig:duty_cycle-b05} we plot AGN duty cycle as a function of Eddington ratio, taking model B05f (left panels) and B05v (right panels) as illustrative examples of the effect of a high $T_C$. We remark that the results for other models are similar. In the top panels, the black and red lines show the cumulative time above and below the given Eddington ratio, respectively. Solid lines are the results compiled from the entire simulation time, while dashed lines only use the data in the final 2 Gyr. The points are observational constraints taken from literature. The black/red points are the measurements of the fraction of objects above/below the given Eddington ratio. The squares, cycles, upward- and downward- pointing triangles are from \citet{Ho2009}, \citet{Greene-Ho2007}, \citet{Kauffmann-Heckman2009} and \citet{Heckman2004}, respectively, which are all compiled from low-redshift observations and should be compared with dashed lines. The star is a constraint compiled from high-redshift observations by \citet{Steidel2003}, which should be compared with solid lines. The top/left panel shows that there are too many AGN burst activities during the entire evolution time in model B05f (solid line), but at the same time the AGN activity in the last 2 Gyr is perhaps too moderate compared to the observational constraints (dashed line). These problems of models B with fixed $T_C$ were already pointed out in \citet{Novak2011}. For this reason it is particularly relevant that in the top/right panel, one can observe a remarkable improvement in model B05v, in which a higher $T_C$ is adopted in the phase of hot accretion flow. This result, combined with the more realistic growth of the black hole we have described in \S\ref{blackholegrowth}, indicates that models B (and their improvements) have good potentialities to be used in realistic simulations.  In fact, the importance of radiative feedback in the kinetic mode can be anticipated from the top/left panel of Fig. \ref{fig:duty_cycle-b05}. We can see from the figure that the galaxy center spends $>80\%$ of its evolution time with Eddington ratios in a range of $0.001-0.02$. Such a range of Eddington ratio belongs to the hot accretion phase (i.e., kinetic mode feedback), according to eq. \ref{eq:model-v}. During this phase, the luminosity is  at most two orders of magnitude lower than the typical quasar mode (assumed to be 0.1 $L_{\rm Edd}$), while the Compton heating efficiency is two orders of magnitude higher in the hot accretion phase than in the cold accretion phases, according to eqs. (\ref{Comptonheating})-(\ref{twotemperature}).

It is well-known that active galaxies spend most of the lifetimes in the radio/quiescent mode, as is shown in the top panels of Fig. \ref{fig:duty_cycle-b05}. By contrast, the Soltan argument indicates that most of the energy is emitted with high Eddington ratio, e.g., an acceptable model must have a significant fraction of the total energy emitted above a level corresponding to $\simeq 10\%L_{Edd}$ \citep[e.g.,][]{Kollmeier2006}. In the bottom panels of Fig. \ref{fig:duty_cycle-b05}, we plot the ``energy-weighted'' duty cycle, the lines are of the same meanings as in the top panels, while they show the fractions of cumulative radiative energy during the evolution time. It is shown by the black solid lines that for model B05v there is 22\% of the total energy emitted with a Eddington ratio $\ge 0.1$, while the value for model B05f is just 3\%. In another word, models ``v'', with a variable Compton temperature, are much more favored than models ``f'' by the Soltan argument.

\subsection{ISM budget: galactic winds and star formation}

The fifth column in Table \ref{tab:Model-Summary} gives the total gas mass $\Delta M_W$
driven beyond 70 kpc (i.e., $\simeq 10$ times the half-light radius of the stellar spheroid).
In this paper, we account the gas driven in these region as the galactic wind.
We can see that the values of $\Delta M_W$ are quite similar for all of the models
despite of the large range of $\epsilon_W^M$ and different treatment of $T_C$. The reason is that AGN feedback by both winds and radiation mainly acts on the few inner kpc, and only in exceptionally strong accretion episodes the resulting shock waves can reach the galaxy outskirts. In the outer region of the galaxy, where most of the galactic winds originate from, AGN feedback does not play an important role and it is mainly SN Ia heating that controls the gas dynamics.

\begin{figure*}[!ht]
\begin{center}
\includegraphics[width=0.90\textwidth]{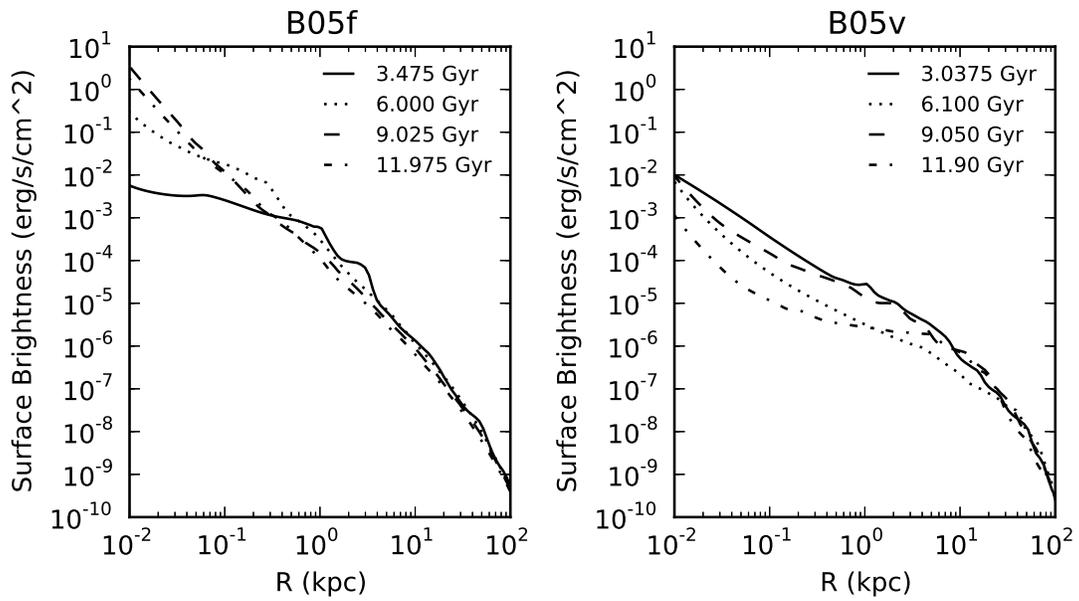}
\end{center}
\caption{X-ray surface brightness profiles of the hot ISM in the band of 0.3-8 kev for models B05f (left panel) and B05v (right panel) at quiescence. The solid, dotted, dashed and dot-dashed lines correspond to increasing times, which are around 3 Gyr, 6 Gyr, 9 Gyr, 12 Gyr, respectively. }
\label{fig:x_ray_properties}
\end{figure*}

The last column in Table \ref{tab:Model-Summary} presents the total ISM mass $M_{\rm gas}$ remaining within 70 kpc at the end of the simulations. The values are similar for all models. This is because it is hard for the radiation and wind to influence such a large spacial region. However, we do find, although not shown here, the mass of the remaining gas in the inner region, say within $\sim 1$ kpc, is significantly smaller in model ``v'' than that in model ``f''. This is again because of the higher $T_C$ thus stronger radiative heating and subsequently lower gas density.  This is similar to the case of the growth of the black hole mass. To examine the effect of a high $T_C$ in a bit more detail, in Figure \ref{fig:x_ray_properties} we plot the X-ray surface brightness profiles (0.3-8 kev; refer to Pellegrini, Ciotti \& Ostriker 2012 for the calculation of the surface brightness) of the hot ISM at quiescence for models B05f (left panel) and B05v (right panel). The solid, dotted, dashed and dot-dashed lines correspond to increasing times, which are around 3 Gyr, 6 Gyr, 9 Gyr, 12 Gyr, respectively. We can see that within $\sim 10$ kpc, the emission profiles become systematically flatter in model ``v'' than in model ``f''.

\begin{figure*}[!ht]
\begin{center}
\includegraphics[width=0.6\textwidth]{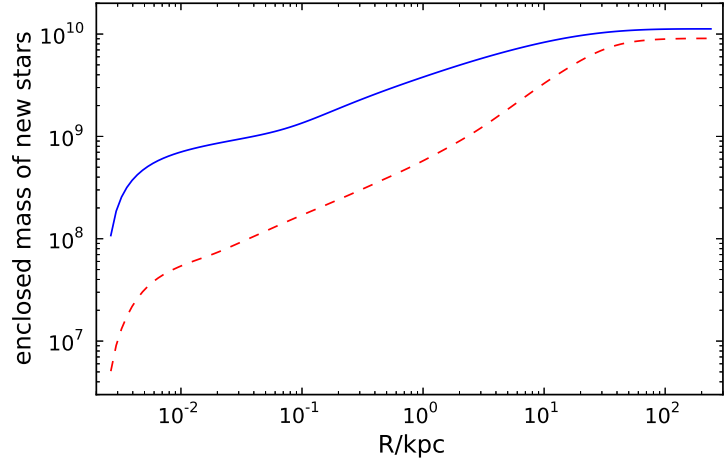}
\end{center}
\caption{Enclosed mass of the newly formed stars at the end of the simulations
(enclosed mass at radius $R$ means the total mass within $R$).
The solid line represent the result of model ``nowind.f'' and dashed line is for model ``nowind.v''.
We can see that the star formation in the galaxy center is heavily suppressed
in model ``nowind.v'' due to the strong radiative heating.}
\label{fig:enclosed-newstar_nowind}
\end{figure*}

Because of the different Compton heating in model ``v'' compared to model ``f'', it is also expected that some difference will be present in the ISM of the two families. In particular, the star formation may also be affected. To examine this effect, we have calculated the total mass of the newly formed stars during the whole evolution epoch. Figure \ref{fig:enclosed-newstar_nowind} shows  the enclosed mass of the newly formed stars as a function of radius $R$ in models ``nowind.v'' and ``nowind.f''. We can see that the star formation in the galaxy center is suppressed in model ``nowind.v'' due to the high Compton temperature by nearly one order of magnitude compared to model ``nowind.f''. However, the star formation in the outer region of the galaxy is just slightly disrupted. This implies that the range of radiative feedback is restricted to be $\la$ 1 kpc.

The suppression of star formation is quite similar to the regulation of black hole growth. In the outer region of the galaxy, the feedback effect on star formation is dominated by supernovae. In the inner region, star formation relates directly to the gas accretion.   The falling cold shell and filament contribute significantly to the mass accretion rate of the black hole, while we usually also find a very high star formation rate there. The strong radiative heating in the hot accretion phase trends to heat the gas into a state of high temperature and low density, which results in the suppression of star formation.

We note that while mechanical feedback by winds could also suppress star formation, the mechanism is different. For mechanical feedback, the gas is expelled outward by ram pressure and probably heated via shocks. But ahead of the shock, gas will be compressed, and thermal instability will grow up rapidly. In this case, mechanical feedback could result in enhanced rather than suppressed star formation. Such a phenomenon is already observed in the numerical simulations by \citet{Liu2013b}. On the other hand, radiative feedback always suppresses star formation.

\subsection{Test run with smoothed Compton temperature}

\begin{figure*}[!ht]
\begin{center}
\includegraphics[width=0.95\textwidth]{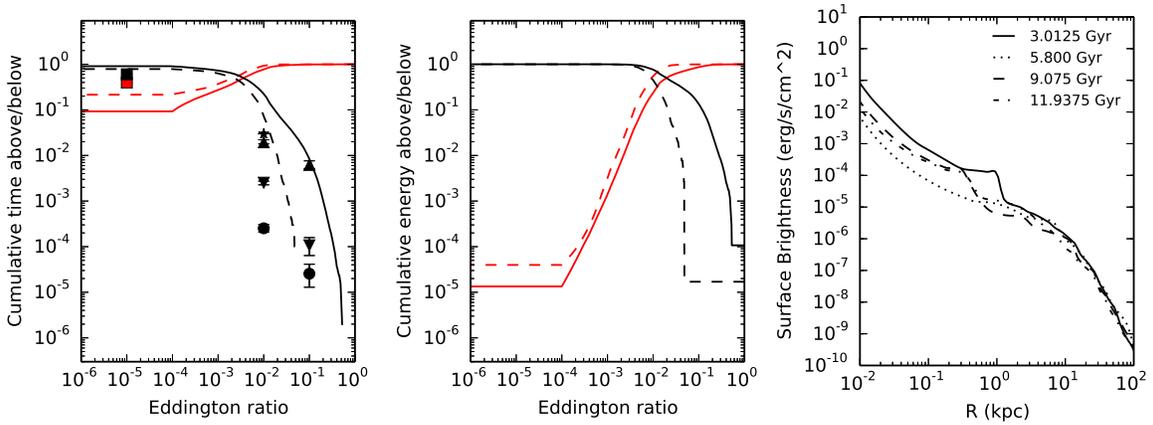}
\end{center}
\caption{Simulation results of model B05s. Right/Middle panel: the AGN duty cycle weighted by time/energy (cf Fig.\ref{fig:duty_cycle-b05}); Left panel: X-ray surface brightness profiles of the hot ISM at quiescence (cf Fig.\ref{fig:x_ray_properties})}
\label{fig:model-b05s}
\end{figure*}

For models ``v'', we use a variable Compton temperature $T_C$ as a piecewise function of the Eddington ratio (see eq. \ref{eq:model-v}). In reality, $T_C$ could change somehow smoothly. In order to mimic such a situation, we design another family of models in which $T_C$ is of smoothed variation, i.e.
\begin{equation}\label{model-s}
T_C = T_{C0}\cdot\frac{1+\dot{m}}{1+A\dot{m}},
\end{equation}
where we choose $T_{C0}=1\times10^9$ K for comparison purpose, and make a test run (denoted as ``B05s'') with the same model parameters as model B05v (see Table \ref{tab:Model-Summary} for the details). We present the simulation results of model B05s in Fig. \ref{fig:model-b05s}. The right and middle panels are the AGN duty cycles weighted by time and energy, respectively (cf Fig. \ref{fig:duty_cycle-b05}); the left panel shows X-ray surface brightness profiles of the hot ISM at quiescence (cf Fig. \ref{fig:x_ray_properties}). We can see that the results are quite similar to those of model B05v, i.e., a smoothed Compton temperature doesn't change our main results.

\section{Discussion and Conclusions}

We have performed two-dimensional high resolution hydrodynamical simulations to investigate the AGN feedback in an isolated elliptical galaxy, focusing in particular on the so far unexposed regime of an accretion-dependent Compton temperature $T_C$. Both radiative and mechanical
feedback are taken into account and also physical processes on galactic scales such as star formation and supernovae are considered in the calculation. The inner boundary of our
computational domain is carefully chosen so as to make sure that the fiducial Bondi radius is resolved in our simulation, allowing for a robust estimate of the black hole accretion rate, which is crucial to study AGN feedback. Following previous works, two types of models have been considered. In ``A'' models, the mechanical efficiency of winds is a constant, while in ``B'' models, it decreases with the decreasing luminosity of AGN as would be expected from radiatively driven winds \citep{Proga2008}. By construction models in the B family are perhaps more realistic than in the A family. However a few important systematic problems have been detected in the B family in previous works. In particular, previous calculations have shown that Model B predicts a too high final black hole mass and incorrect AGN duty cycle compared to observations. A major improvement of the present work compared with previous ones is that we allow for a different Compton temperature of the radiation spectrum from the central AGN. In the radiation mode, the accretion flow is described as a standard thin disk thus the corresponding Compton temperature is relatively low, $T_C\simeq 10^7$K. But in the kinetic mode, the accretion flow is described by a hot accretion flow thus the corresponding Compton temperature should be higher. Following previous theoretical work, we adopt $T_C\simeq 10^9$K. This in general results in a stronger radiative feedback as adopted in \citet{Ciotti2001}. Consequently, we find that the accretion processes are more chaotic and further suppressed. Specifically, we find that Model B now predicts a correct range of black hole mass and AGN duty cycle. In more general sense, our study indicate that in the kinetic mode, radiative feedback is much more important than previous thought and should be seriously considered in future studies.

Some improvements can be made based on the present work in the future. First, as we have mentioned in the paper, we have simply chosen $T_C \simeq 10^9$ K as the Compton temperature of the emitted spectrum from the hot accretion flows. This value comes from the calculation based on the theoretical spectrum from a hot accretion flow. In reality, the accretion flow is composed of an inner hot accretion flow plus  an outer truncated thin disk (see \citeauthor{Yuan-Narayan2014} \citeyear{Yuan-Narayan2014} for a review). The best way is to combine the observed spectral energy distribution of LLAGNs with various luminosities and calculate the Compton temperature as a function of luminosity. Moreover, relativistic effect may be important in calculating $T_C$ which was unfortunately neglected in the previous work.

A complementary aspect of this study that needs an improvement is the description of the properties of the nuclear wind as a function of the black hole accretion state. In particular, issues line the mechanical efficiency $\epsilon_W$, the wind speed and its angular distribution. In fact, as we have shown in this paper, $\epsilon_W$ is of crucial importance for the evaluation of mechanical feedback. But unfortunately this value is still poorly constrained by observations or by theories so we have to treat it as a free parameter, although some initial results have been obtained, e.g., in the study of hot accretion flows \citep{Yuan2012b,Li2013}. The angular distribution of wind, which  is also expected to be important to mechanical feedback, is again poorly constrained. In this paper we assume a ``bipolar-like'' angular distribution function of
the wind. In the future, it is expected that the theoretical studies will be able to better constrain these two properties. In fact, we are using the MHD numerical simulation data to study the wind from hot accretion flows and some initial results have been obtained. For example, we find that the winds do have a wide range of distribution angles, from $\simeq 0-45^{\circ}$. But different from the ``bipolar-like'' structure, we find that most of the mass flux of wind seem to be concentrate on the disk surface. It will be interesting to examine its effect in cooperation with the variations of $T_C$.


\section{Acknowledgments}
We thank Scott Tremaine for useful discussion of the Soltan argument, and Silvia Pellegrini for the comments of the X-ray properties of the ISM, and the referee for the helpful comments and suggestions. This work was supported in part by the Natural Science Foundation of China (grants 11103061, 11133005 and 11121062), the National Basic Research Program of China (973 Program, grant 2014CB845800), and the Strategic Priority Research Program ``The Emergence of Cosmological Structures'' of the Chinese Academy of Sciences (grant XDB09000000). L.C. is supported by the grant PRIN MIUR 2010-2011 ``The chemical and dynamical evolution of the Milky Way and Local Group galaxie'' 2010LY5N27. G.S.N. was supported by the ERC European Research Council under the Advanced Grant Program Num 267399 Momentum. The simulations were carried out at Shanghai Supercomputer Center and SHAO Super Computing Platform.



\begin{table*}[ht]

\caption{Some parameters and calculation results of the computed models}
\label{tab:Model-Summary}

\begin{center}
\begin{tabular}{lcccccc}
\hline
Model & $\epsilon_W^M$      & $\left< \epsilon_{EM} \right>$ & $\log \Delta M_{BH}$
      & $\log \Delta M_W$ & $\log M_{gas}$
\tabularnewline
\hline

A0f   &$5.0\times10^{-3}$& 0.03731 & 7.2678 & 9.8027 & 9.6923   \tabularnewline
A05f  &$1.0\times10^{-3}$& 0.05597 & 8.0671 & 9.7337 & 9.6416   \tabularnewline
A1f   &$2.5\times10^{-4}$& 0.06644 & 8.7466 & 9.7155 & 9.6152   \tabularnewline
A2f   &$1.0\times10^{-4}$& 0.07204 & 9.1099 & 9.8545 & 9.5876   \tabularnewline
A3f   &$5.0\times10^{-5}$& 0.07330 & 9.5489 & 9.8541 & 9.6570   \tabularnewline

$\cdots$ & $\cdots$ & $\cdots$ & $\cdots$ & $\cdots$ & $\cdots$ \tabularnewline

B0f   &$5.0\times10^{-3}$& 0.06040 & 9.3773 & 9.8173 & 9.5989   \tabularnewline
B05f  &$1.0\times10^{-3}$& 0.06515 & 9.9414 & 9.8904 & 9.6141   \tabularnewline
B1f   &$2.5\times10^{-4}$& 0.06781 & 10.006 & 9.8974 & 9.5763   \tabularnewline
B2f   &$1.0\times10^{-4}$& 0.06678 & 10.004 & 9.8839 & 9.5796   \tabularnewline
B3f   &$5.0\times10^{-5}$& 0.06644 & 10.001 & 9.8857 & 9.5788   \tabularnewline
B4f   &$3.0\times10^{-5}$& 0.06635 & 10.001 & 9.8839 & 9.5810   \tabularnewline

$\cdots$ & $\cdots$ & $\cdots$ & $\cdots$ & $\cdots$ & $\cdots$ \tabularnewline
nowind.f  & ---          & 0.06611 & 9.995  & 9.8864 & 9.5867   \tabularnewline

\hline \hline
A0v   &$5.0\times10^{-3}$& 0.03688 & 7.2411 & 9.8636 & 9.6431   \tabularnewline
A05v  &$1.0\times10^{-3}$& 0.06753 & 8.0549 & 9.7355 & 9.6569   \tabularnewline
A1v   &$2.5\times10^{-4}$& 0.07037 & 8.6848 & 9.7593 & 9.5612   \tabularnewline
A2v   &$1.0\times10^{-4}$& 0.07199 & 9.0242 & 9.8312 & 9.5776   \tabularnewline
A3v   &$5.0\times10^{-5}$& 0.07408 & 9.2183 & 9.9005 & 9.6118   \tabularnewline

$\cdots$ & $\cdots$ & $\cdots$ & $\cdots$ & $\cdots$ & $\cdots$ \tabularnewline

B0v   &$5.0\times10^{-3}$& 0.06094 & 8.8918 & 9.7935 & 9.6323   \tabularnewline
B05v  &$1.0\times10^{-3}$& 0.06729 & 9.2255 & 9.8228 & 9.6153   \tabularnewline
B1v   &$2.5\times10^{-4}$& 0.07032 & 9.3446 & 9.8228 & 9.6203   \tabularnewline
B2v   &$1.0\times10^{-4}$& 0.07072 & 9.4096 & 9.7659 & 9.6058   \tabularnewline
B3v   &$5.0\times10^{-5}$& 0.07116 & 9.4290 & 9.7257 & 9.6259   \tabularnewline
B4v   &$3.0\times10^{-5}$& 0.07349 & 9.4160 & 9.8104 & 9.5830   \tabularnewline

$\cdots$ & $\cdots$ & $\cdots$ & $\cdots$ & $\cdots$ & $\cdots$ \tabularnewline
nowind.v  & ---          & 0.07392 & 9.4580 & 9.7951 & 9.6704   \tabularnewline

\hline\hline
B05s  &$1.0\times10^{-3}$& 0.06461 & 9.3854 & 9.8103 & 9.5856   \tabularnewline

\hline
\end{tabular}
\end{center}

Notes: (1) Parameter $\epsilon_W^M$ is the mechanical wind efficiency, $\left< \epsilon_{EM} \right>$ is the mass-weighted mean radiative efficiency of the black hole accretion flow, $ \Delta M_{BH}$ is the change in the mass of the black hole in the evolution, $\Delta M_W$ is the total gas mass driven beyond 70 kpc (which is far from the center and the ISM there is recognized as galactic wind in this paper) and $M_{gas}$ is the total mass remaining within 70 kpc. All the masses are in unit of solar mass; (2) There is no mechanical feedback of disk wind in models ``nowind.f'' and ``nowind.v''.
\end{table*}

\begin{thebibliography}{}


\bibitem[Arav et al.(2005)]{Arav2005}
Arav, N., Kaastra, J., Kriss, G. A., et al.\ 2005, \apj, 620, 665


\bibitem[Binney(2001)]{Binney2001}
Binney, J. \ 2001, APSC, 250, 481


\bibitem[Blustin et al.(2007)]{Blustin2007}
Blustin, A.~J. et al.\ 2007, \aap, 466, 107


\bibitem[Bondi(1952)]{Bondi1952}
Bondi, H.\ 1952, \mnras, 112, 195


\bibitem[Brighenti \& Mathews(2002)]{Brighenti2002}
Brighenti, F., Mathews, W. G., \ 2002, \apj, 573, 542

\bibitem[Brighenti \& Mathews(2003)]{Brighenti2003}
Brighenti, F., Mathews, W. G., \ 2003, \apj, 587, 580


\bibitem[Cattaneo \& Best(2009)]{Cattaneo-Best2009}
Cattaneo, A. \& Best, P.~N.\ 2009, \mnras, 395, 518


\bibitem[Chartas et al.(2003)]{Chartas2003}
Chartas, G., Brandt, W.~N. \& Gallagher, S.~C.\ 2003, \apj, 595, 85


\bibitem[Chiaberge et al.(2006)]{Chiaberge2006}
Chiaberge, M., Gilli, R., Macchetto, F. D., Sparks, William B. 2006, \apj, 651, 728


\bibitem[Ciotti et al.(1991)]{Ciotti1991}
Ciotti, L., D'Ercole, A., Pellegrini, S. \& Renzini, A.\ 1991, \apj, 376, 380


\bibitem[Ciotti \& Ostriker(1997)]{Ciotti1997}
Ciotti, L. \& Ostriker, J.~P.\ 1997, \apjl, 487, L105


\bibitem[Ciotti \& Ostriker(2001)]{Ciotti2001}
Ciotti, L. \& Ostriker, J.~P.\ 2001, \apj, 551, 131


\bibitem[Ciotti \& Ostriker(2007)]{Ciotti2007}
Ciotti, L. \& Ostriker, J.~P.\ 2007, \apj, 665, 1038


\bibitem[Ciotti et al.(2009a)]{Ciotti2009a}
Ciotti, L., Morganti, L. \& de Zeeuw, P.~T.\ 2009a, \mnras, 393, 491


\bibitem[Ciotti et al.(2009b)]{Ciotti2009b}
Ciotti, L., Ostriker, J.~P. \& Proga, D.\ 2009b, \apj, 699, 89


\bibitem[Ciotti et al.(2010)]{Ciotti2010}
Ciotti, L., Ostriker, J.~P. \& Proga, D.\ 2010, \apj, 717, 708


\bibitem[Ciotti \& Ostriker(2012)]{Ciotti2012}
Ciotti, L. \& Ostriker, J.~P.\ 2012, in {\it Hot Interstellar Matter in Elliptical Galaxies},
Kim D.-W. \& Pellegrini S.(eds.), ASSL, 378, 83


\bibitem[Crenshaw et al.(2003)]{Crenshaw2003}
Crenshaw, D.~M., Kraemer, S.~B. \& George, I.~M.\ 2003, \araa, 41, 117


\bibitem[Crenshaw \& Kraemer(2012)]{Crenshaw2012}
Crenshaw, D. M., Kraemer, S. B. 2012, \apj, 753, 75


\bibitem[Czoske et al.(2008)]{Czoske2008}
Czoske, O., Barnab{\`e}, M., Koopmans, L.~V.~E., Treu, T. \& Bolton, A.~S.\ 2008, \mnras, 384, 987


\bibitem[Dai et al.(2008)]{Dai2008}
Dai, X., Shankar, F. \& Sivakoff, G.~R.\ 2008, \apj, 672, 108


\bibitem[Di Matteo et al.(2005)]{DiMatteo2005}
Di Matteo, T., Springel, V., Hernquist, L. 2005, \nat, 433, 604


\bibitem[Djorgovski \& Davis(1987)]{Djorgovski1987}
Djorgovski, S. \& Davis, M.\ 1987, \apj, 313, 59


\bibitem[Done et al.(2007)]{Done2007}
Done, C., Gierlinski, M., Kubota, A. 2007, Astron. Astrophys. Rev, 15, 1


\bibitem[Dressler et al.(1987)]{Dressler1987}
Dressler, A., Lynden-Bell, D., Burstein, D., et al.\ 1987, \apj, 313, 42


\bibitem[Dye et al.(2008)]{Dye2008}
Dye, S., Evans, N.~W., Belokurov, V., Warren, S.~J. \& Hewett, P.\ 2008, \mnras, 388, 384


\bibitem[Eracleous et al.(2010)]{Eracleous2010}
Eracleous, M., Hwang, J. A., Flohic, H. M. L. G. 2010 \apjs, 187, 135


\bibitem[Faber \& Jackson(1976)]{Faber1976}
Faber, S.~M. \& Jackson, R.~E.\ 1976, \apj, 204, 668


\bibitem[Fabian(2012)]{Fabian2012}
Fabian, A.~C.\ 2012, \araa, 50, 455


\bibitem[Fender et al.(2003)]{Fender2003}
Fender, R. P., Gallo, E., Jonker, P. G. 2003, \mnras, 343, 99


\bibitem[Ferrarese \& Merritt(2000)]{Ferrarese2000}
Ferrarese, L. \& Merritt, D.\ 2000, \apjl, 539, L9


\bibitem[Gaspari et al.(2012)]{Gaspari2012}
Gaspari, M., Brighenti, F. \& Temi, P.\ 2012, \mnras, 424, 190


\bibitem[Gaspari et al.(2013)]{Gaspari2013}
Gaspari, M., Ruszkowski, M. \& Oh, S. Peng \ 2013, \mnras, 432, 3401


\bibitem[Gebhardt et al.(2000)]{Gebhardt2000}
Gebhardt, K. et al.\ 2000, \apjl, 539, L13


\bibitem[Gibson et al.(2009)]{Gibson2009}
Gibson, R.~R. et al.\ 2009, \apj, 692, 758


\bibitem[Graham et al.(2011)]{Graham2011}
Graham, A.~W., Onken, C.~A., Athanassoula, E., \& Combes, F.\ 2011, \mnras, 412, 2211


\bibitem[Greene \& Ho(2007)]{Greene-Ho2007}
Greene, J. E. \& Ho, L. C. 2007, \apj, 667, 131


\bibitem[G{\"u}ltekin et al.(2009)]{Gueltekin2009}
G{\"u}ltekin, K. et al.\ 2009, \apj, 698, 198


\bibitem[H{\"a}ring \& Rix(2004)]{Haering2004}
H{\"a}ring, N. \& Rix, H.-W.\ 2004, \apjl, 604, L89


\bibitem[Haiman et al.(2004)]{Haiman2004}
Haiman, Z., Ciotti, L. \& Ostriker, J.~P.\ 2004, \apj, 606, 763


\bibitem[Hamann et al.(2008)]{Hamann2008}
Hamann, F., Kaplan, K.~F., Rodr{\'{\i}}guez Hidalgo, P., Prochaska, J.~X.
\& Herbert-Fort, S.\ 2008, \mnras, 391, L39


\bibitem[Hayes et al.(2006)]{Hayes2006}
Hayes, J.~C., Norman, M.~L., Fiedler, R.~A., et al.\ 2006, \apjs, 165, 188


\bibitem[Heckman et al.(2004)]{Heckman2004}
Heckman, T. M., Kauffmann, G., Brinchmann, J. et al. 2004, \apj, 613, 109


\bibitem[Ho(1999)]{Ho1999}
Ho L. C. 1999, \apj, 516, 672


\bibitem[Ho(2008)]{Ho2008}
Ho L. C., 2008, \araa, 46, 475


\bibitem[Ho(2009)]{Ho2009}
Ho L. C., 2009, \apj, 699, 626


\bibitem[Jiang et al.(2010)]{Jiang2010}
Jiang, Y.-F., Ciotti, L., Ostriker, J. P., Spitkovsky, A. 2010, \apj, 711, 125


\bibitem[Johansson et al.(2009)]{Johansson2009}
Johansson, P. H., Naab, T., Burkert, A. 2009, \apj, 690, 802


\bibitem[Kalemci et al.(2013)]{Kalemci2013}
Kalemci, E., Dincer, T., Tomsick, J. A., 2013, \apj, 779, 95


\bibitem[Kauffmann \& Heckman(2009)]{Kauffmann-Heckman2009}
Kauffmann, G. \& Heckman, T. M. 2009, \mnras, 397, 135


\bibitem[Kollmeier et al.(2006)]{Kollmeier2006}
Kollmeier, Juna A., Onken, Christopher A., Kochanek, Christopher S. et al. 2006, \apj, 648, 128


\bibitem[Kormendy \& Richstone(1995)]{Kormendy1995}
Kormendy, J. \& Richstone, D.\ 1995, \araa, 33, 581


\bibitem[Kormendy \& Ho(2013)]{Kormendy2013}
Kormendy, J., Ho, L.~C.\ 2013. \araa, 51, 511


\bibitem[Kurosawa et al.(2009)]{Kurosawa2009}
Kurosawa, R., Proga, D., Nagamine, K. 2009, \apj, 707, 823


\bibitem[Kurosawa \& Proga(2009a)]{Kurosawa-Proga2009a}
Kurosawa, R. \& Proga, D.\ 2009a, \mnras, 397, 1791


\bibitem[Kurosawa \& Proga(2009b)]{Kurosawa-Proga2009b}
Kurosawa, R. \& Proga, D.\ 2009b, \apj, 693, 1929


\bibitem[Kulier et al.(2013)]{Kulier2013}
Kulier, A., Ostriker, J. P., Natarajan, P., Lackner, C. N., Cen, R. 2013, arXiv:1307.3684v1


\bibitem[Li et al.(2013)]{Li2013}
Li, J., Ostriker, J., Sunyaev, R. 2013, \apj, 767, 105


\bibitem[Liu et al.(2013a)]{Liu2013a}
Liu, C., Yuan, F., Ostriker, J. P., Gan, Z., Yang, X.\ 2013a, \mnras, 434, 1721


\bibitem[Liu et al.(2013b)]{Liu2013b}
Liu, C., Gan, Z., Xie, F.\ 2013b, RAA, 13, 899


\bibitem[Magorrian et al.(1998)]{Magorrian1998}
Magorrian, J. et al.\ 1998, \aj, 115, 2285


\bibitem[Maraston(2005)]{Maraston2005}
Maraston, C.\ 2005, \mnras, 362, 799


\bibitem[Marconi \& Hunt(2003)]{Marconi2003}
Marconi, A. \& Hunt, L.~K.\ 2003, \apjl, 589, L21


\bibitem[Mathews \& Brighenti(2003)]{Mathews2003}
Mathews, W. G., Brighenti, F., \ 2003, \araa, 41, 191


\bibitem[McNamara \& Nulsen(2007)]{McNamara2007}
McNamara, B.~R. \& Nulsen, P.~E.~J.\ 2007, \araa, 45, 117


\bibitem[Narayan \& Yi(1994)]{Narayan1994}
Narayan, R. \& Yi, I.\ 1994, \apjl, 428, L13


\bibitem[Narayan \& Yi(1995)]{Narayan-Yi1995}
Narayan, R. \& Yi, I.\ 1995, \apj, 452, 710


\bibitem[Narayan et al.(2012)]{Narayan2012}
Narayan, R., Sadowski, A., Penna, R. F., Kulkarni, A. K. 2012, \mnras, 426, 3241


\bibitem[Novak et al.(2011)]{Novak2011}
Novak, G.~S., Ostriker, J.~P. \& Ciotti, L.\ 2011, \apj, 737, 26


\bibitem[Novak et al.(2012)]{Novak2012}
Novak, G.~S., Ostriker, J.~P. \& Ciotti, L.\ 2012, \mnras, 427, 2734


\bibitem[Ostriker \& Ciotti(2005)]{Ostriker2005}
Ostriker, J.~P. \& Cotti, L. \ 2005, Phil. Trans. Royal Soc. A, 363, 667


\bibitem[Ostriker et al.(2010)]{Ostriker2010}
Ostriker, J.~P., Choi, E., Ciotti, L., Novak, G.~S. \& Proga, D.\ 2010, \apj, 722, 642


\bibitem[Park \& Ostriker(1999)]{Park-Ostriker1999}
Park, M.-G. \& Ostriker, J.~P.\ 1999, \apj, 527, 247


\bibitem[Park \& Ostriker(2001)]{Park-Ostriker2001}
Park, M.-G. \& Ostriker, J.~P.\ 2001, \apj, 549, 100


\bibitem[Park \& Ostriker(2007)]{Park2007}
Park, M.-G. \& Ostriker, J.~P.\ 2007, \apj, 655, 88


\bibitem[Parriott \& Bregman(2008)]{Parriott2008}
Parriott, J.~R., \& Bregman, J.~N.\ 2008, \apj, 681, 1215


\bibitem[Pellegrini(2010)]{Pellegrini2010}
Pellegrini, S.\ 2010, \apj, 717, 640

\bibitem[Pellegrini et al.(2012)]{Pellegrini2012a}
Pellegrini, S., Ciotti, L. \& Ostriker, J.~P.\ 2012, \apj, 744, 21


\bibitem[Pellegrini(2012)]{Pellegrini2012}
Pellegrini, S.\ 2012, in {\it Hot Interstellar Matter in Elliptical Galaxies},
Kim D.-W. \& Pellegrini S.(eds.), ASSL, 378, 21


\bibitem[Peterson \& Fabian(2006)]{Peterson2006}
Peterson, J.~R. \& Fabian, A.~C.\ 2006, \physrep, 427, 1


\bibitem[Proga \& Kallman(2004)]{Proga2004}
Proga, D. \& Kallman, T.~R.\ 2004, \apj, 616, 688


\bibitem[Proga et al.(2008)]{Proga2008}
Proga, D., Ostriker, J.~P. \& Kurosawa, R.\ 2008, \apj, 676, 101


\bibitem[Rusin \& Kochanek(2005)]{Rusin2005}
Rusin, D. \& Kochanek, C.~S.\ 2005, \apj, 623, 666


\bibitem[Sadowski et al.(2013)]{Sadowski2013}
Sadowski, A., Narayan, R., Penna, R., Zhu, Y.\ 2013, \mnras, 436, 3856


\bibitem[Sazonov et al.(2004)]{Sazonov2004}
Sazonov, S.~Y., Ostriker, J.~P. \& Sunyaev, R.~A.\ 2004, \mnras, 347, 144


\bibitem[Sazonov et al.(2005)]{Sazonov2005}
Sazonov, S.~Y., Ostriker, J.~P., Ciotti, L. \& Sunyaev, R.~A.\ 2005, \mnras, 358, 168


\bibitem[Shin et al.(2010)]{Shin2010}
Shin, M.-S., Ostriker, J.~P. \& Ciotti, L.\ 2010, \apj, 711, 268


\bibitem[Soltan(1982)]{Soltan1982}
Soltan, A. \ 1982, \mnras, 200, 115

\bibitem[Springel et al.(2005)]{Springel2005}
Springel, V., Di Matteo, T. \& Hernquist, L.\ 2005, \mnras, 361, 776


\bibitem[Steidel et al.(2003)]{Steidel2003}
Steidel, C. C., Adelberger, K. L., Shapley, A. E. et al. 2003, \apj, 592, 728


\bibitem[Stone \& Norman(1992)]{Stone1992}
Stone, J.~M. \& Norman, M.~L.\ 1992, \apjs, 80, 753


\bibitem[Tombesi et al.(2013)]{Tombesi2013}
Tombesi, F., Cappi, M., Reeves, J.~N. et al.\ 2013, \mnras, 430, 1102


\bibitem[Tremaine et al.(2002)]{Tremaine2002}
Tremaine, S. et al.\ 2002, \apj, 574, 740


\bibitem[Wang et al.(2013)]{Wang2013}
Wang, Q. D., Nowak, M. A., Markoff, S. B., et al. 2013, Science, 341, 981


\bibitem[Xie \& Yuan(2012)]{Xie-Yuan2012}
Xie, F. \&  Yuan, F.\ 2012, \mnras, 427, 1580


\bibitem[Younes et al.(2010)]{Younes2010}
Younes, G., Porquet, D., Sabra, B., Reeves, J. N., Grosso, N. 2010, A\&A, 539, 104


\bibitem[Yu \& Tremaine(2002)]{Yu2002}
Yu, Q. \& Tremaine, S.\ 2002, \mnras, 335, 965

\bibitem[Yu, Yuan \& Ho(2011)]{Yu2011}
Yu, Z., Yuan, F. \& Ho, L.~C.\ 2011, \apj, 726, 87

\bibitem[Yuan et al.(2009)]{YXO2009}
Yuan, F., Xie, F. \& Ostriker, J.~P.\ 2009, \apj, 691, 98


\bibitem[Yuan \& Li(2011)]{Yuan-Li2011}
Yuan, F., Li, M. 2011, \apj, 737, 23


\bibitem[Yuan et al.(2012)]{Yuan2012b}
Yuan, F., Bu, D. \& Wu, M.\ 2012, \apj, 761, 130

\bibitem[Yuan \& Narayan(2014)]{Yuan-Narayan2014}
Yuan, F., Narayan, R. 2014, ARA\&A, in press (arXiv:1401.0586)


\end{thebibliography}
\end{document}